\documentclass[journal]{IEEEtranTIE}
\usepackage{graphicx}
\usepackage{cite}
\usepackage{picinpar}
\usepackage{amsmath}
\usepackage{url}
\usepackage[latin1]{inputenc}
\usepackage{colortbl}
\usepackage{soul}
\usepackage{multirow}
\usepackage{pifont}
\usepackage{color}
\usepackage{alltt}
\usepackage{hyperref}
\usepackage{enumerate}
\usepackage{siunitx}
\usepackage{breakurl}
\usepackage{epstopdf}
\usepackage{pbox}
\usepackage{verbatim}
\usepackage{amssymb}
\usepackage{algorithm,algorithmic}
\usepackage{amsthm}
\usepackage{cleveref}

\usepackage{amsfonts}
\usepackage{textcomp}
\usepackage{xcolor}
\usepackage{float}
\usepackage{caption}
\usepackage{subcaption}

\newtheorem{definition}{Definition}
\newtheorem{theorem}{Theorem}

\usepackage[normalem]{ulem}

\renewcommand{\baselinestretch}{0.88}

\usepackage[absolute]{textpos}
\begin{document}
\makeatletter
\def\ps@IEEEtitlepagestyle{%
	\def\@oddfoot{\mycopyrightnotice}%
	\def\@evenfoot{}%
}
\def\mycopyrightnotice{
	{\footnotesize This work has been submitted to the IEEE for possible publication. Copyright may be transferred without notice,
		after which this version may no longer be accessible. \hfill}
	\gdef\mycopyrightnotice{}
}

\title{Adaptive control of a mechatronic system using constrained residual reinforcement learning}
\author{
	\vspace*{-8pt}
	{Tom Staessens, Tom Lefebvre
    and Guillaume Crevecoeur}

	\thanks{
	This research received funding from the Flemish Government under the "Onderzoeksprogramma Artificiële Intelligentie (AI) Vlaanderen" programme.

	T. Staessens, T. Lefebvre and G. Crevecoeur	are with the Department of Electrical Energy, Metals, Mechanical Constructions and Systems, Ghent University, 9000 Ghent, Belgium, and also with EEDT-DC, Flanders Make (e-mail: \{tom.staessens, tom.lefebvre, guillaume.crevecoeur\}@ugent.be).

}
\vspace*{-24pt}} 

\maketitle

\begin{abstract}
We propose a simple, practical and intuitive approach to improve the performance of a conventional controller in uncertain environments using deep reinforcement learning while maintaining safe operation. Our approach is motivated by the observation that conventional controllers in industrial motion control value robustness over adaptivity to deal with different operating conditions and are suboptimal as a consequence. Reinforcement learning on the other hand can optimize a control signal directly from input-output data and thus adapt to operational conditions, but lacks safety guarantees, impeding its use in industrial environments. To realize adaptive control using reinforcement learning in such conditions, we follow a residual learning methodology, where a reinforcement learning algorithm learns corrective adaptations to a base controller's output to increase optimality. We investigate how constraining the residual agent's actions enables to leverage the base controller's robustness to guarantee safe operation. We detail the algorithmic design and propose to constrain the residual actions relative to the base controller to increase the method's robustness. Building on Lyapunov stability theory, we prove stability for a broad class of mechatronic closed-loop systems. We validate our method experimentally on a slider-crank setup and investigate how the constraints affect the safety during learning and optimality after convergence.

\end{abstract}

\begin{IEEEkeywords}
mechatronics, servo systems, motion control, reinforcement learning, uncertain systems.
\end{IEEEkeywords}

\markboth{}%
{}

\vspace*{-6pt}
\section{Introduction}

\IEEEPARstart{M}{echatronic} drivetrain systems face increasingly demanding performance and efficiency requirements in industrial and manufacturing applications. They furthermore need to become more autonomous while interacting with a varying environment. Adequate motion control needs to address these challenges. When designing, implementing and tuning controllers, having knowledge on the dynamics of the mechatronic system is key. Accurate position and speed control of servo drive systems for instance require detailed knowledge on the inertia and friction \cite{kim2018moment}. Based on ab initio physical modelling principles it is possible to approximate the real system behavior. However capturing the full mechatronic system dynamics is often cumbersome and challenging as mechatronic systems are plagued by nonlinear and complex dynamic behavior due to interacting components \cite{kim2018moment, wang2019adaptive, papageorgiou2018robust}. Parameter identification procedures can subsequently be engaged to closer align the model to the real mechatronic system \cite{schon2011system}. Nonetheless, despite tremendous engineering modeling efforts, uncertainties may still be present. 

Unfortunately the optimality of control system design is strongly affected by the modelling fidelity \cite{mayne2000constrained}. When designed off-line, controllers are approximate due to the inherent uncertainties in the real-world that are not incorporated in the modelling. They furthermore require tremendous tuning efforts, e.g. finding gains in PID controllers to track setpoints. A wide range of adaptive control strategies have been designed to alleviate this issue on-line: adapting PID gains \cite{kuc2000adaptive}, online gravity compensation \cite{yang2021new} or, starting from an approximate model, adapting the parameters in linearly parameterized model predictive control (MPC) \cite{adetola2011robust} or of a fuzzy approximation of the remaining unknown system influences in sliding-mode control \cite{yang2021adaptive}. In the stochastic optimal control framework, stochastic uncertainties are introduced to the deterministic optimal control resulting in linear quadratic Gaussian \cite{athans1971role} and stochastic model predictive control formalisms \cite{mesbah2016stochastic}, that are variants of the LQR and deterministic MPC, respectively. In the framework of MPC, off-line design strategies include open-loop and feedback minimax formulations \cite{lofberg2003minimax} and tube-based formulations. Initially, robust tube-based MPC has been used for linear systems in process control \cite{limon2010robust} and more recently has been further elaborated for mechatronic systems \cite{yan2016tube}. Tube-based MPC approaches the control problem by first solving the MPC problem for the nominal system. In a second stage the ancillary state feedback control law is designed to confine the error between nominal and actual states within an invariant tube.

Strategies such as the tube-based MPC are able to cope with uncertainties but do not aim to reduce them by learning intelligently from the input-output behavior of the controlled system. Learning-based techniques with iterative learning control techniques \cite{wang2009survey} have been devised that adapt with respect to the tracking error. These iterative learning control algorithms can work under a model-free assumption \cite{chi2017improved}, but can only compensate for periodic disturbances. Also to adapt for repetitive tasks, a recent MPC strategy was proposed relying solely on a data-driven model \cite{rosolia2018data}. Reinforcement learning (RL) methods on the other hand adapt to the actual behavior of the mechatronic system, interacting with a variable environment, and directly learn an optimal feedback policy (referred to as the agent). As opposed to the aformentioned techniques, the learned policy is state-dependent, where state can be interpreted broadly as any measurement information, e.g. camera images, and is independent of any time-, state- or trajectory-dependent periodicity or underlying dynamics.

Hence RL opens up interesting new perspectives on adaptivity. Assuming the parametrized policy is sufficiently expressive, the training procedure of RL is capable of generalizing to variable operating conditions and changing environmental settings \cite{lewis2009reinforcement}. Moreover, RL can handle control problems that are difficult to approach with conventional controllers because the control goal can be specified indirectly as a term in a reward function with no explicit requirements on its form or dependencies. The main disadvantage of RL is that it can only fashion new insights by interacting with the system and the environment in real-time, leveraging the actions it takes to probe and explore the optimality landscape. Since the former process relies on the stochasticity inherent to the system or on deliberate perturbation, this may lead to unsafe situations limiting the usage of RL to non safety critical situations \cite{dulac2019challenges}. Recently, related work for the adaptive and robust control of nonlinear systems using actor critic RL has been developed. These methods however still require an approximate model or system identification, are designed for reference tracking only \cite{fu2020mrac, na2020adaptive} or rely on the convergence of the critic function approximators and as such the quality of the collected dataset, for robustness guarantees. \cite{radac2020robust}.

Next to aforementioned stability issues, the amount of explorative trials to learn and find optimal control actions is significant. To face the issue of data-efficient training the literature proposes the use of off-policy algorithms  \cite{wang2016sample}. As opposed to on-policy algorithms \cite{schulman2017proximal}, off-policy methods train the agent based on data generated with another, canonical controller. As a consequence, off-policy algorithms strongly improve the data-efficiency, however they still require a significant amount of trials which may not always be feasible for real-world applications \cite{dulac2019challenges}. Nevertheless, the simple observation that off-policy algorithms can be merged with conventional controllers is of particular interest and sheds a new light on the adaptivity issue raised earlier.

In this paper we explore the possibilities of a residual architecture to cater the limitations bothering the straightforward application of traditional RL in an industrial setting. On the one hand we will rely on a traditional suboptimal but stabilising control law. On the other hand we superpose a residual RL agent that may adapt the control output in an attempt to optimize an auxiliary objective. This architecture, coined residual reinforcement learning (RRL), has been explored in earlier research and results into an efficient, safe and optimal control design. RRL has been introduced recently to alleviate the exploration needs and increase tractability in terms of data-efficiency for data-driven robot control \cite{johannink2019residual, silver2018residual}. By applying the reinforcement learning algorithm residually on a base controller that roughly approaches the control objective, the base controller `guides' the reinforcement learning algorithm to an approximate solution, accelerating training. 
The constraints imposed on the residual agent were absolute, determined by the limits of the controlled system's inputs.

Originally used to accelerate learning in robot control, RRL can be engaged to increase the optimality in mechatronic motion control. The situation differs however for the setting of industrial motion control of mechatronic systems, where the challenge shifts to maintaining safe operation while realising an adaptive motion control.
By leveraging a base controller, residual RL improves the data-efficiency that prohibits traditional RL from being used in such real-world mechatronic systems. However the safety concerns, being the main reason impeding its in industrial settings, remain unaddressed.

This paper explores the possibilities of adding RL next to an existing control law in a safe and robust manner. We do this by following a residual approach for which we introduce relative constraints on the residual agent. The ensuing objective of this paper is to realize an adaptive motion control for mechatronic systems using RL, applicable in a real-world setting. This is approached in a twofold manner. First, we detail the design of a stable constrained residual learning methodology. We introduce an algorithmic adaptation to residual RL, employing relative constraints and prove the stability of both methods using the Lyapunov method. Second, we validate the method's ability to achieve adaptive control, improving the performance for motion control of a mechatronic system compared to the traditional controller. We provide implementation details on the presented methodology and demonstrate the results by applying it on a slider crank setup and evaluating the Mean Absolute Error (MAE) of the objective.
The contributions of this paper are as follows:
\begin{itemize}
	\item Extension of the Residual Reinforcement Learning framework to Constrained RRL. This allows the use of RL algorithms in industrial, safety-critical settings and as such enables online adaptive control without any assumptions or prerequisites of the system dynamics, control objective or form of the controller inputs.
	\item Theoretical analysis of the developed method using Lyapunov stability theory, proving stability for a broad class of mechatronic systems even under worst case conditions.
	\item Experimental validation on a slider-crank, a non-linear system with applications in many industrial systems.
\end{itemize}
In conclusion, where pure RL does not work, we show that this framework enables the use of these algorithms in real-world mechatronic settings.

\section{Methodology}

\subsection{Reinforcement Learning}

Reinforcement learning operates within a standard (Partially Observable) Markov Decision Process (MDP) framework. 
An MDP is a tuple $M = (\mathcal{S}, \mathcal{A}, r, p)$ where $s \; \in \; \mathcal{S}$ are states, $a \; \in \; \mathcal{A}$ are actions, 
$r(s, a)$ is the reward for taking action $a$ in state $s$ and $p(s_{t+1}\vert s_t, a_t)$  is the probability of transitioning to state $s_{t+1}$ following state $s_t$ and action $a_t$. We define a trajectory $\tau$ as a sequence of states and actions, $\tau = (s_0, a_0, s_1, a_1, ...)$. The return is defined as the infinite discounted sum of rewards $R(\tau) = \lim_{T\rightarrow \infty}\sum_{t=0}^{T} \gamma^t r(s_t,a_t)$ with $0 \leqslant \gamma \leqslant 1$ the temporal discount factor. Given an initial state $s_0$, the objective of any RL method is to solve the following stochastic optimal control problem by finding an optimal policy $\pi^*(a_t \vert s_t)$ that maximizes the expected return
\begin{equation}
\begin{aligned}
J(\pi) &= \mathbb{E}_{\tau\sim p(\tau|\pi)}[R(\tau)] \\
\pi^* &= \arg\!\max_\pi J(\pi)
\end{aligned}
\end{equation}
with 
$p(\tau \vert s_0, \pi) = \Pi_{t=0}^{T-1} p(s_{t+1} \vert s_t,a_t) \pi(a_t \vert s_t)$. In this paper, the Soft Actor-Critic (SAC) algorithm is employed as RL method for all experiments. SAC is a state-of-the-art actor-critic RL method. This implies that both estimates of the state-action value function and policy are approximated using a neural network. Based on temporal differencing these estimates are iterated until they satisfy the Bellman equation. SAC is unique with regard to other actor-critic methods as it maintains a stochastic actor. Actions are realized by sampling from a Gaussian distribution whose mean and variance are outputted by the network. This has the advantage of encouraging exploration during training and achieving a higher stability after convergence. For further details we refer to \cite{haarnoja2018soft, haarnoja2018softalgandapp}.
\vspace{-0.2cm}
\subsection{Constrained Residual Reinforcement Learning}

Recently introduced for robot control, residual reinforcement learning trains an RL controller residually on top of an imperfect, traditional controller \cite{johannink2019residual, silver2018residual}. The RL algorithm leverages the traditional controller as an initialization to enable data-efficient reinforcement learning for tasks where traditional RL is intractable, such as robotic insertion tasks where rewards are sparse \cite{schoettler2019deep}. Starting from a suboptimal, but adequate and robust controller, as often present in the motion control of industrial applications, we introduce the Constrained Residual Reinforcement Learning (CRRL) architecture.

Two advantages are principle to the concept of CRRL. Firstly, the architecture leverages the traditional controller to guarantee robust exploration of the RL agent during operation. The robust controller can be tuned so that the exploration of the residual policy remains within the principle region of attraction. Secondly, as with basic RRL, the traditional controller provides a good initialization for the reinforcement learning algorithm which may further improve the steepness of the learning curve.

In this contribution we study two variants of the basic RRL architecture, absolute and relative CRRL.

\subsubsection{Absolute CRRL}
For Absolute CRRL, we simply superpose the residual policy, parametrized by the parameters $\theta$, to the traditional controller. This produces a control input $u_{\theta}(s)$
\begin{equation}
\label{eq:absresidual}
  u_{\theta}(s) = u(s) + \beta_{a}\pi_{\theta}(y(s))    
\end{equation}
\noindent where $u(s)$ is the traditional control algorithm, $\pi_{\theta}(y(s))$ is the reinforcement learning policy, $y(s)$ is a preprocessing feature extraction map and $\beta_{a}$ is a parameter determining the scale of the residual actions. Besides the state $s$, the map $y(s)$ can contain e.g. $u(s)$ or any other feature that seems interesting\footnote{Generally speaking the feature map may even contain additional measurement information such as camera images etc.}. Note that (\ref{eq:absresidual}) corresponds mathematically with basic RRL employing a RL algorithm which uses a $\tanh$ to confine its actions, such as SAC \cite{haarnoja2018soft}, with special care taken to tune $\beta_a$.

In robot control, e.g. connector insertion tasks \cite{schoettler2019deep}, choosing $\beta_{a}$ as an absolute constraint so as to confine the total actions within the feasible action space, such as the torque limits of the controlled actuators, suffices to learn a tractable policy that succeeds in its task. For industrial motion control however, safety during all phases of the learning process is required. Determining bounds for the residual agent that guarantee safety is non-trivial. In Section \ref{sec:lyapunov}, we establish safety conditions that guarantee that regardless of the residual agent the system stays within the principle region of attraction of the traditional controller. However the detailed modelling requirements may make it unfeasible for some real-world applications in the industry. Alternatively, for cyclic processes, one can take the output of the base controller during one cycle as a reference. This can however not be done in a straightforward manner when facing non-cyclic processes. Furthermore, it requires the scale of the base actions to remain consistent within one cycle to provide a safe constraint tube.

\subsubsection{Relative CRRL}
\begin{figure}[t!]
	\centerline{\includegraphics[width=.7\columnwidth]{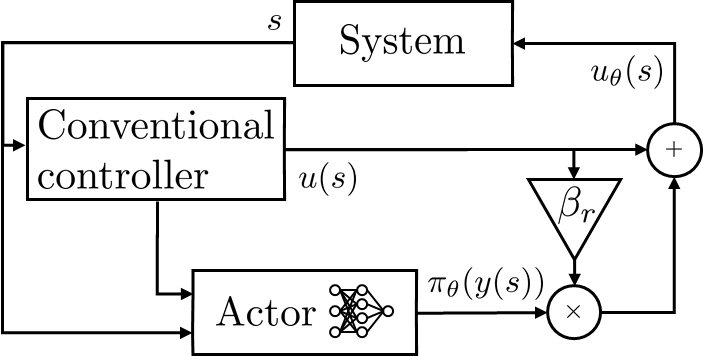}}
	\caption{Overview of the proposed relative CRRL structure.}
	\label{fig:overview}
	\vspace*{-0.5cm}
\end{figure}

To realize a residual agent maintaining safe operation irrespective of such considerations regarding the underlying process, we propose to constrain its actions relative to the base controller's outputs. By constraining the reinforcement learning algorithm to a percentage of the classical control action and adding it to said action, we effectively create a tube around the classical control actions where the reinforcement learning algorithm is allowed to explore and learn corrective adaptations to the conventional controller's output that improve its performance. Both during the learning phase, when exploration is the dominant behavior, and when encountering inputs that deviate from the training distribution after convergence, the classical control algorithm determines the bulk of the control input and thereby ensures a safe and robust operation. The resulting relative residual policy is:
\begin{equation}
\label{eq:relresidual}
  u_{\theta}(s) = u(s) (1 + \beta_r\pi_{\theta}(y(s))   ) 
\end{equation}
\noindent with $\beta_r > 0$ a parameter constraining the actions of the neural network relative to the actions of the base control algorithm. Fig. \ref{fig:overview} shows an overview of the relative CRRL structure.

\subsubsection{Algorithm} The implementation of the absolute CRRL architecture is straightforward. That of the relative CRRL architecture is more subtle. As the actor network is trained through gradient descent to minimize its loss \cite{haarnoja2018soft}, multiplying the actor output by $\beta_r u(s)$ directly scales the gradient of all actor network parameters by this fraction of the base controller's output. This gives more weight to situations where the base controller's output is large during the training of the actor network. To alleviate this imbalance, one can opt to use the unscaled output for training and only scale the action when applying it on the system itself. The input state for the networks can then be extended with the base controller's output to again have full state information. However, in practice, we have found the residual controller's performance to benefit from the scaling during training as well. Therefore the former option is used throughout the experiments.
\subsubsection{Convergence} 
	\label{sec:convergence}
	For the convergence condition, we rely on the proof given in \cite{haarnoja2018soft} for SAC in the theoretical tabular case, which is approximated for practical use in continuous domains by using the neural networks as function approximators. By considering the base controller as a part of the system on which the residual SAC agent acts and assuming its robustness, the same conditions for convergence hold. The adapted algorithm of \cite{haarnoja2018softalgandapp}, which automatically balances the stochasticity of the actor as a function of the reward, further promotes convergence in practice by lowering the variance of the policy in states where the policy achieves high rewards.
\vspace{-0.2cm}
\subsection{Closed-loop tracking stability guarantees for mechanical systems using a PD base-line controller}
\label{sec:lyapunov}
Here we analyse the closed-loop stability of the proposed learning approach. We consider a CRRL controller both with absolute (\ref{eq:absresidual}) and relative constraints (\ref{eq:relresidual}). The experimental performance of both controllers is investigated further in Section \ref{section:constraints}. Our synthesis focusses on closed-loop tracking stability guarantees for mechanical systems using a PI base-line controller $f(s)$. Such provides a generic setting that meets the requirements of many practical examples from industry. We aim to establish safety guarantees when worst-case conditions are met during exploration, not to make claims about its near optimal behaviour during convergence. Therefore, we treat the residual agent as a disturbance whose actions destabilize the system. Our analysis allows us to determine robust gain values for the base-line controller that guarantee stability regardless of the actions taken by the residual agent. 

Our analysis is based on classic Lyapunov stability theory and can be summarized in the following two theorems. We note that the proofs of both theorems rely on a particular choice of Lyapunov function. Therefore, these theorems are illustrative to the fact that it is possible to obtain conditions for the base-line control settings corresponding specific safety guarantees in the context of CRRL. On the other hand, our conditions might be overly conservative and possibly weaker conditions exist based on other Lyapunov functions.

\begin{theorem}
	\label{th:absolute}
Consider a mechanical system with generalised coordinates $q\in\mathbb{R}^n$, input $\tau\in\mathbb{R}^n$ and reference trajectory $q_r \in \mathbb{R}^n$ so that $\|\dot{q}_r\|\leq \omega_0$ and $\|\ddot{q}_r\|\leq \alpha_0$ whose dynamics are governed by the equation
\vspace*{-.16cm}
\begin{equation}
{M}(q)\ddot{q} + B(q,\dot{q})\dot{q} + g(q) = u
\end{equation}
With $e=q-q_r$ let the \underline{absolute} CRRL policy be defined as 
\begin{equation}
u(q,\dot{q}) = - k_\text{P}\dot{e} -k_{\text{I}}  e - \beta_a\pi_\theta(q,\dot{q},\dots )
\end{equation}
Then closed-loop error trajectories $x=(e,\dot{e})$ are bounded by
\begin{equation}
\|x(t)\| \leq e^{-\frac{1}{2} \frac{\alpha_3}{\alpha_2} t} \frac{\alpha_2}{\sqrt{\alpha_1}} \|x(0)\|^2+\delta, ~ \lim_{t\rightarrow \infty} \|x(t)\| \leq \delta
\end{equation}
with $\alpha_1$, $\alpha_2$, $\alpha_3$ and $\alpha_4$ defined as in theorem \ref{th:quad} and where
\begin{equation}
\delta = \tfrac{\mu_\eta(P)\sqrt{\tfrac{1}{4}k_\text{P}^2 + \lambda^2 \mu_\eta(M)^2} \left(\alpha_0 + \tfrac{1}{\nu_\eta(M)}\left(\mu_\eta(B)\omega_0 + g_0 + \beta_a \right)\right)}{\nu_\eta(P)\nu_\eta(\check{Q})}
\end{equation}
where $g_0 = \sup \|g(q)\|$ 
and matrices $P$ and $\check{Q}$ defined as in theorem \ref{th:quad} if also $
\lambda > \frac{k_\text{P}^2}{2k_\text{I}\nu(M)}$, 
$k_{\text{P}} > \frac{2\mu_\eta(L_q)\omega_0}{1-\frac{1}{\lambda}-\sqrt{\frac{\epsilon^2\mu_\eta(P)\mu_\eta(B)}{\lambda^2 \nu_\eta(P)\nu_\eta(M)}}}$ and 
$k_{\text{I}} > \frac{\mu_\eta(M)\mu_\eta(B)^2 \omega_0^2}{2\nu(M)^2}$.
\end{theorem}
\noindent For the proof, definitions of the norms we refer to appx. \ref{sec:proof-of-theorem-refthabsolute}.

\begin{theorem}
		\label{th:relative}
	Considering the same controlled system as in theorem \ref{th:absolute}. Let the \underline{relative} CRRL policy be defined as
	\begin{equation}
	u(q,\dot{q}) = -\left(k_{\text{P}} \dot{e} + k_{\text{I}} e \right) (1 + \beta_r\pi_\theta(q,\dot{q},\dots ) )
	\end{equation}
	Then closed-loop error trajectories $x$ are bounded by
	\begin{equation}
	\|x(t)\| \leq e^{-\frac{1}{2} \frac{\alpha_3}{\alpha_2} t} \frac{\alpha_2}{\sqrt{\alpha_1}} \|x(0)\|^2+\delta, ~ \lim_{t\rightarrow \infty} \|x(t)\| \leq \delta
	\end{equation}
	\vspace*{-.5cm}
	where
	\begin{equation}
	\delta = \tfrac{\mu_\eta(P)\sqrt{\tfrac{1}{4}k_\text{P}^2 + \lambda^2 \mu_\eta(M)^2} \left(\alpha_0 + \tfrac{1}{\nu_\eta(M)}\left(\mu_\eta(B)\omega_0 + g_0 \right)\right)}{\nu_\eta(P)\nu_\eta(\check{Q})}
	\end{equation}
	if it also holds that
	$\lambda > \frac{k_\text{P}^2}{2(1-2\beta_r )k_\text{I}\nu(M)}$,\\ 
	$k_{\text{P}} > \frac{2 \mu_\eta(L_q)\omega_0}{1-\frac{1}{\lambda}-\frac{2\beta_r k_{\text{I}}}{\lambda}-\sqrt{\frac{\epsilon^2\mu_\eta(P)\mu_\eta(B)}{\lambda^2 \nu_\eta(P)\nu_\eta(M)}}}$ and  
	$k_{\text{I}} > \frac{\mu_\eta(M)\mu_\eta(B)^2 \omega_0^2}{2(1-\beta_r)\nu(M)^2}$.

\end{theorem}
\noindent For the proof we refer to appx. \ref{sec:proof-of-theorem-refthabsolute}.

These two theorems suggest that if the closed-loop system is initiated in a state $\|x(0)\|=0$, the error will never grow beyond a magnitude $\delta$. Provided a desired value for $\delta$ and depending on the CRRL architecture, it is possible to choose values for $k_\text{P}$, $k_\text{I}$ and $\lambda$ and therefore for $\beta_a$ or $\beta_r$, so that the conditions in either theorem \ref{th:absolute} or \ref{th:relative} are satisfied. For practical calculation, note that all variables in Theorems \ref{th:absolute} and \ref{th:relative} depend only on the norms of the matrices, defined in appx. \ref{sec:proof-of-theorem-refthabsolute}. These norms can be calculated with knowledge of the initial state bounds as per the proof's assumptions. For the slider crank system defined in Table \ref{tab:sc_details} and a controller with parameters $k_p=1.1$ and $k_i=2.3$, the conditions for $\beta_r$ are 0.36, 35.28 and 0.999 respectively to ensure stable convergence.
\vspace*{-.2cm}
\section{Results and Discussion} 
In this contribution we study the CRRL methodology on a physical slider-crank setup. A PI controller is chosen to obtain a stable system. Combining RL with more complex controllers such as MPC controllers for online parameter tuning \cite{zanon2020safe} or state-feedback controllers is possible, but is out of the scope of this paper: to improve upon and be directly modular with a commonly used controller in an industrial setting. Note that the CRRL framework allows for the use of any base controller nonetheless.
\begin{figure}[htbp]
	\centerline{\includegraphics[width=\columnwidth]{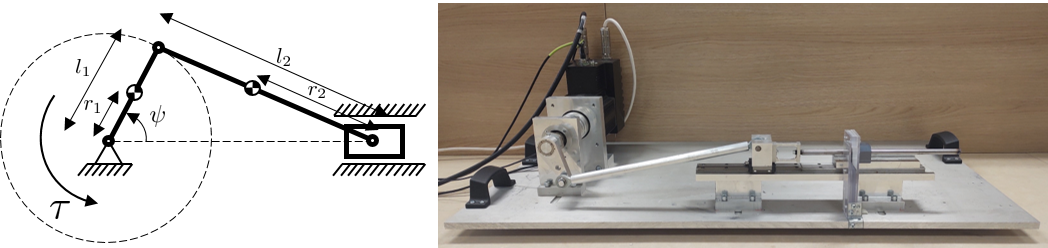}}
	\caption{Schematic overview of a slider-crank \cite{de2019neural} and picture of the experimental setup.}
	\vspace*{-.5cm}
	\label{fig:sc}
\end{figure}
\subsection{Experimental setup}

\subsubsection{Slider-crank linkage}
Present in many industrial applications, a slider-crank provides reciprocating linear motion through a rotary motor in combination with a bar linkage system. Fig. \ref{fig:sc} shows a schematic overview as well as a picture of the experimental setup. The dimensions of the setup are detailed in Table \ref{tab:sc_details}. This system exhibits highly nonlinear behavior \cite{de2019neural} and is often plagued by unidentified load disturbances and unknown interactions with the environment. To achieve an adequate control under these conditions, typically some form of PID controller is used. This strategy is adequate in applications with low requirements on precision, but suffers from suboptimality for systems with varying loads or environment conditions. Coping with these uncertainties requires either re-tuning the controller or having knowledge of the disturbances interacting with the system which is often unfeasible in practice. This application is of direct relevance in various industrial systems, e.g. compressors \cite{gao2010filter},  hydraulic  pumps  \cite{li2019design},  weaving  looms  \cite{eren2005comparison}  and  presses \cite{zheng2014modeling}.

\begin{table}[htbp]
	\caption{Slider-crank setup details.}
	\vspace*{-0.3cm}
	\begin{center}
		\begin{tabular}{l|r}
			\hline
			Parameter & Value \\
			\hline
			Length of link 1 ($l_1$) & $0.05 \; m$ \\
			Length of link 2 ($l_2$) & $0.275 \; m$ \\
			Distance to link 1 center of mass ($r_1$) & $0.33 \; m$ \\
			Distance to link 2 center of mass ($r_2$) & $0.1375 \; m$ \\
			Inertia of link 1 & $0.0038 \; kg \cdot m^2$ \\
			Inertia of link 2 & $0.002193  \; kg \cdot m^2$ \\
			Mass of link 1 & $0.223 \; kg$ \\
			Mass of link 2 & $0.348 \; kg$ \\
			Mass of sliding block & $0.795 \; kg$ \\
			Motor friction coefficient & $0.0047$ \\
			\hline
		\end{tabular}
		\label{tab:sc_details}
	\end{center}
	\vspace*{-0.5cm}
\end{table}

\subsection{Experiments}

The control problem that we consider is tracking of an angular speed reference through torque control of the system. Due to the nonlinear behavior of the system and large influence of friction on its dynamics, this requires a nonlinear control signal that is sensitive to external influences, making it a challenging task for a PID controller to achieve high performance. All results shown in this section are averaged over at least five runs with mean and min-max shown. The SAC policy, which is implemented customly to allow for the algorithmic changes,  is trained using the mean squared error (MSE) of the instantaneously measured deviation from the required rotational velocity, $\frac{1}{2}(\omega_d - \omega)^2$. In the discussion of the results, the Mean Absolute Error (MAE) is used to allow to interpret the algorithm's properties intuitively as this focuses less on the outliers. In the figures however, both metrics are shown.

The sampling of the action is kept on for the entire experiment here, since the stochasticity is limited after convergence as described in Section \ref{sec:convergence}. The state used is composed of the crank's rotational velocity and the sine and cosine of its angle, requiring no extra sensors but the encoder standardly used for the base controller. As with any RL algorithm, tuning of the hyperparameters is a necessary step to obtain the desired performance for CRRL as well. Nonetheless, we have found the robustness to hyperparameter settings of SAC with automatic temperature adjustment \cite{haarnoja2018softalgandapp} to hold for the CRRL employing SAC as well, with only the batch size and learning rate having a notable effect on the outcome, provided no unconventional values for the other parameters are set. To illustrate this robustness and to allow easy comparison, the same SAC hyperparameters, listed in Table \ref{tab:sac_hyperpars}, are used throughout all CRRL experiments in this paper. The computation time for 1 epoch on a computer with a 6-core Intel i7-8700 CPU with 8GB RAM is 0.87 seconds. Note that this is only necessary for training the network. During deployment, only a forward pass of the actor network at each timestep is needed, which consists of only 2436 FLOPS. Section \ref{section:learning_process} discusses the general behavior of a CRRL controller followed by a more in-depth examination of its different features in the subsequent subsections.\\

\begin{table}[htbp]
	\vspace*{-0.5cm}
	\caption{SAC parameters.}
	\vspace*{-0.3cm}
	\begin{center}
		\begin{tabular}{l|r|r}
			\hline
			Parameter & CRRL & RL \\
			\hline
			Optimizer (all networks) & Adam & Adam \\
			Learning rate (all networks) & 3e-4 & 1e-5 \\
			Discount $(\gamma)$ & 0.97 & 0.9 \\
			Batch size (randomly sampled from replay buffer) & 256 & 256 \\
			Replay buffer size & 1e6 & 1e6 \\
			Number of hidden layers - Actor network & 2 & 2 \\
			Number of hidden layers - Critic network & 3 & 2 \\
			Number of neurons per hidden layer - Actor network & 32 & 32 \\
			Number of neurons per hidden layer - Critic network & 128 & 32 \\
			Nonlinearity & ReLU & ReLU \\
			Target smoothing coefficient & 0.005 & 0.005 \\
			\hline
		\end{tabular}
		\label{tab:sac_hyperpars}
		\vspace*{-0.3cm}
	\end{center}
\end{table}

\subsubsection{Learning process}
\label{section:learning_process}

Fig. \ref{fig:perf_example} shows the performance of a relative CRRL controller with $\beta_r=0.2$ and an averagely well tuned PI controller ($k_p = 1.4$, $k_I = 0.1$) as base policy tracking a constant angular reference of 60 rpm. This PI controller, as well as the optimally and poorly tuned ones employed later have been tuned through a grid search on the system for each reference signal. The grid ranges from 0.1 to 1.2 and 2.6 for $k_i$ and $k_p$ respectively in steps of 0.1. These bounds are determined empirically. $\omega_d$ denotes the desired and $\omega$ the actual crank angular velocity. The general form of the learning process displayed by this configuration is illustrative for all other configurations mentioned hereafter. 

The performance of a PI controller on the slider crank setup varies slightly from run to run despite lab conditions with limited external disturbances. Therefore each experiment starts with an initial run-in phase in which only the PI controller acts on the system to benchmark the results (the blue shaded region in Fig. \ref{fig:perf_example}). This variability of the PI controller's performance over different runs is illustrative for the difficulties in optimally tuning a controller for all conditions. Note that the reward shown is not the instantaneous reward of one timestep, but the average reward over one revolution. As such, the error offset of the PI controller does not illustrate a steady-state error, but the inability of the PI controller to compensate for the non-linearities throughout one revolution which amounts to the mean error shown. After this phase, the residual controller is activated at epoch 65. For our experiments, an epoch is defined as 500 measurement points sent to the PC. In the beginning of training, the Q-values are small and the entropy term dominates in the objective \cite{haarnoja2018soft}. This leads to actions that are sampled nearly uniformly from the distribution output by the SAC policy \cite{haarnoja2018soft}, resulting in random and therefore possibly unsafe controller outputs. In Fig. 1 one can see that  the drop in performance caused by the exploration phase in CRRL is, although unavoidably still present, strongly limited. This intuitively corresponds to assuming a robustness of the base controller to limited disturbances, i.e. the residual actions constrained by the parameter $\beta_r$, as can often be assumed of an industrially employed controller and confirms the theoretical findings of Section \ref{sec:lyapunov}. The effect of constraining the residual actions to the base policy is examined further in Section \ref{section:constraints}. After the exploration phase, the performance of the RL algorithm improves until it converges to a residual policy that gives a stable improvement of approximately 13\% compared to the base PI controller.

\subsubsection{RL benchmark} 
\begin{figure}[t]
	\vspace*{-0.3cm}
	\centerline{\includegraphics[width=.9\columnwidth]{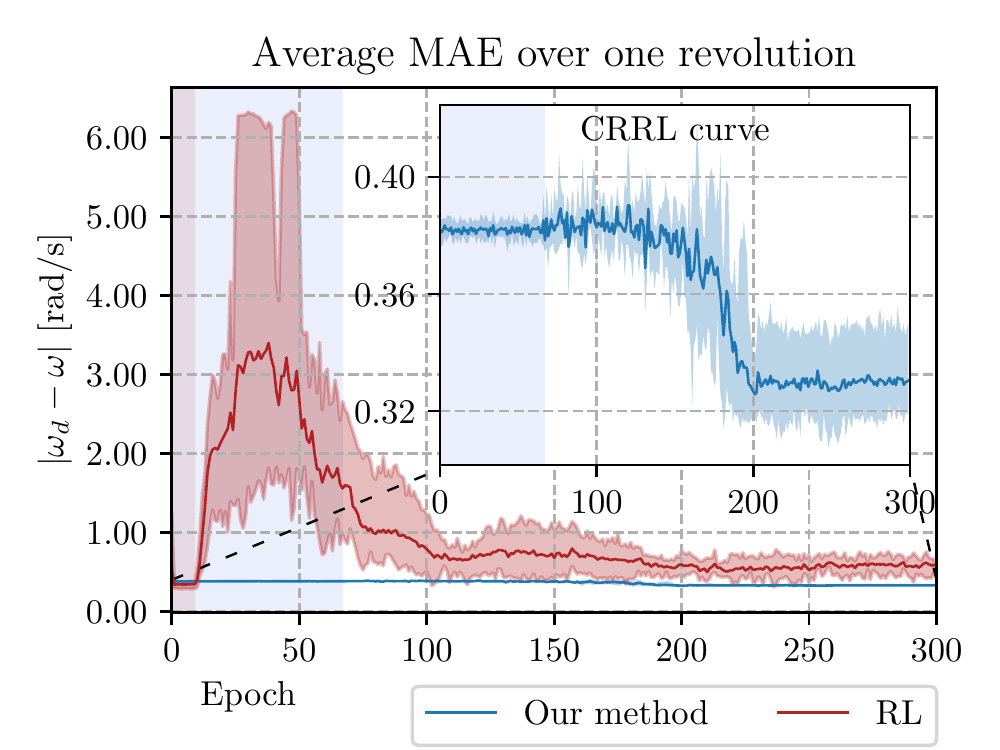}}
	\caption{CRRL and pretrained RL controller loss, averaged over 5 runs with min and max shown. During the blue and red shaded region respectively, only the base controller acts.}
	\label{fig:perf_example}
	\vspace*{-0.5cm}
\end{figure}

A standalone SAC controller was pretrained to mimic an optimally tuned base controller and subsequently trained to further optimize the learning objective. A suitable hyparparameter combination, listed in Table \ref{tab:sac_hyperpars}, was obtained after a two week period of manual tuning. The resulting loss curve is shown on Fig. \ref{fig:perf_example} with the red shaded region indicating the pretraining period. For this best performing set of parameters, the algorithm converges to a MAE of approximately 0.55 rad/s after 200 epochs. During exploration, the error unavoidably reaches up to 6.28 rad/s occasionally, i.e. standstill, due to the full freedom of the SAC algorithm. This indicates that the residual controller benefits from the base controller both for limiting the unsafe behaviour during exploration and converging to a high performing policy. Having reached the limit of performance possible by tuning the base algorithm, it also demonstrates the modularity of CRRL to existing controllers without needing system specific adaptations such as policy constraints, specific reward shaping ... which would be required to further decrease the standalone RL algorithm's error.

\subsubsection{The importance of constraining residual actions}
\label{section:constraints}

\begin{figure}[h!]
	\vspace*{-0.3cm}
	\centering
	\begin{subfigure}{.45\columnwidth}
		\centering
		\includegraphics[width=\linewidth]{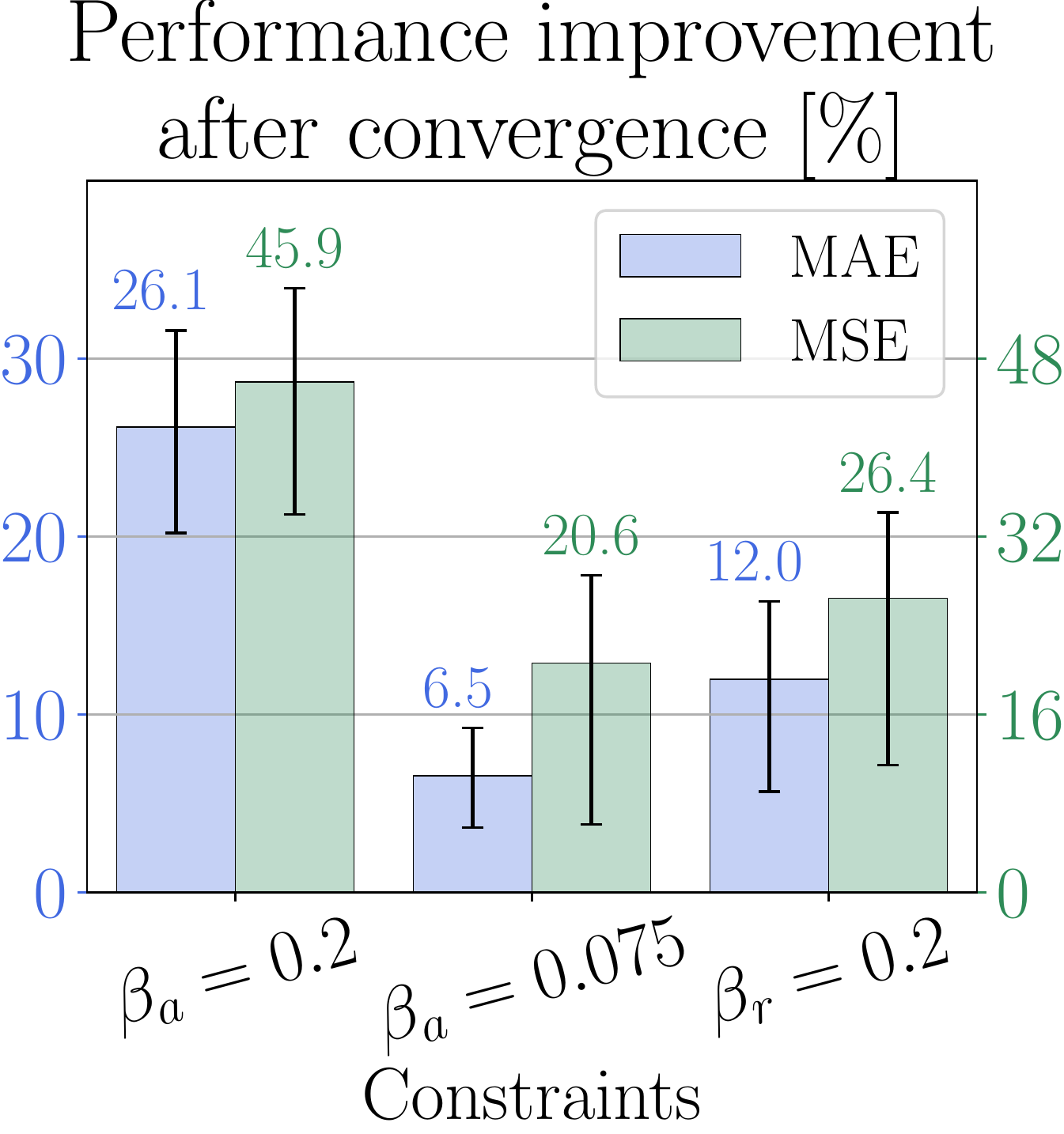}
		\vspace*{-.3cm}
		\label{fig:improv_relabs}
	\end{subfigure}%
	\begin{subfigure}{.45\columnwidth}
		\centering
		\includegraphics[width=\linewidth]{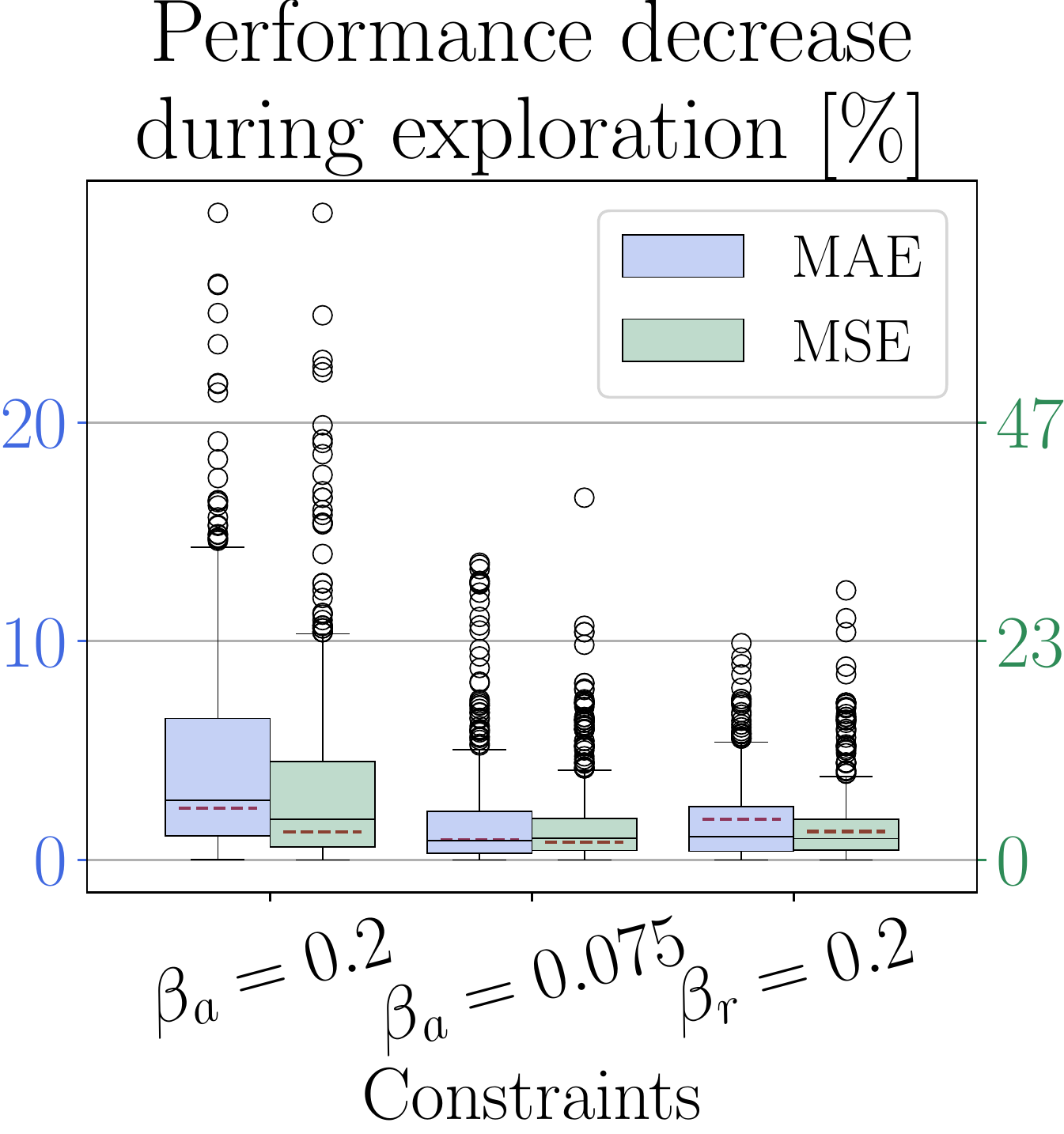}
		\vspace*{-.3cm}
		\label{fig:error_relabs}
	\end{subfigure}
	\caption{Comparison of the performance after convergence and during exploration for a CRRL controller with different constraints.}
	\vspace*{-0.3cm}
	\label{fig:rel_vs_abs}
\end{figure}

In relative CRRL (\ref{eq:relresidual}), the actions are constrained to a tube with width a percentage relative to the base controller's output. Fig. \ref{fig:rel_vs_abs} compares CRRL with relative constraints to a residual policy with absolute constraints (\ref{eq:absresidual}), tracking a constant angular velocity reference of 60 rpm. To ease the comparison, the bounds of the absolute tube are expressed as a percentage of the largest base controller output during a cycle without residual controller. The base controller is an averagely tuned PI controller ($k_p = 1.4$, $k_I = 0.1$). Fig. \ref{fig:rel_vs_abs} on the left shows the average performance improvement after convergence of the CRRL controller relative to the average PI controller performance during the first 65 epochs. The right side shows a boxplot of the relative decrease in performance of all epochs after activating the residual controller where the reward was lower than the average PID reward. The dotted red line indicates the largest negative deviation by the PID controller itself from its average reward during the first 65 epochs. These conventions are maintained for the remainder of the results.

For a residual controller within an absolute tube of 20\%, both the decrease in performance during exploration and the improvement after convergence are substantially larger than for a relative tube of 20\%, as is to be expected due to the increased freedom given to the residual controller. In the next paragraph, this trade-off is discussed in more detail. To compare with the improvement obtained by a relative bound, we experimentally found a residual controller with an absolute tube of 7.5\% to have a similar decrease in performance as a relatively constrained controller of 20\%. The final increase in performance however reaches only approximately 54\% of the increase reached by a relatively constrained controller. This indicates that the relative constraint method of (\ref{eq:relresidual}) is advantageous to achieve a higher optimality while maintaining safe operation. For all experiments throughout the remainder of this paper, this method is employed.\\ 

\begin{figure}[h!]
	\centering
	\begin{subfigure}{.45\columnwidth}
		\centering
		\includegraphics[width=\linewidth]{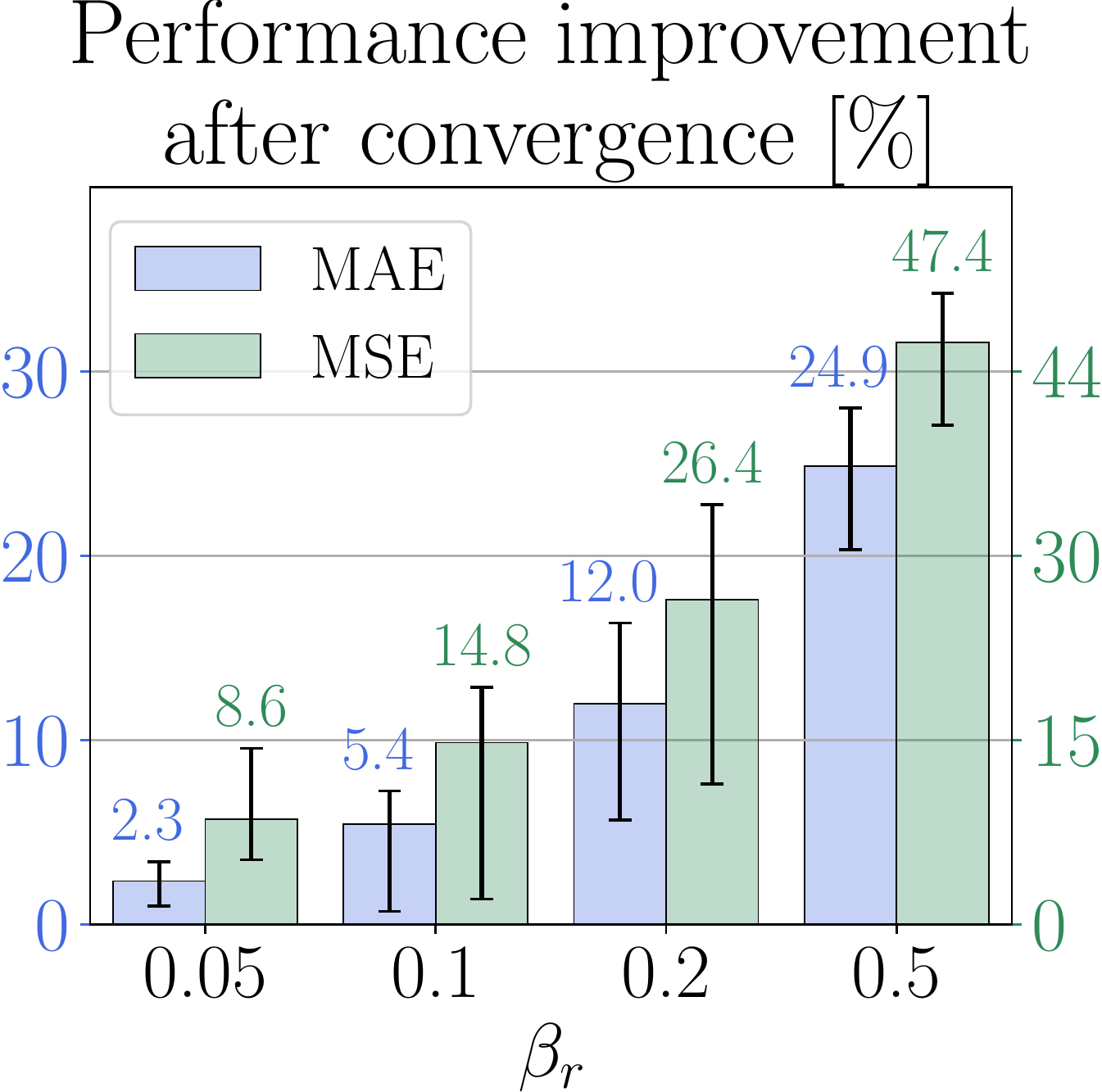}
		\vspace*{-.3cm}
		\label{fig:improv_relT}
	\end{subfigure}%
	\begin{subfigure}{.45\columnwidth}
		\centering
		\includegraphics[width=\linewidth]{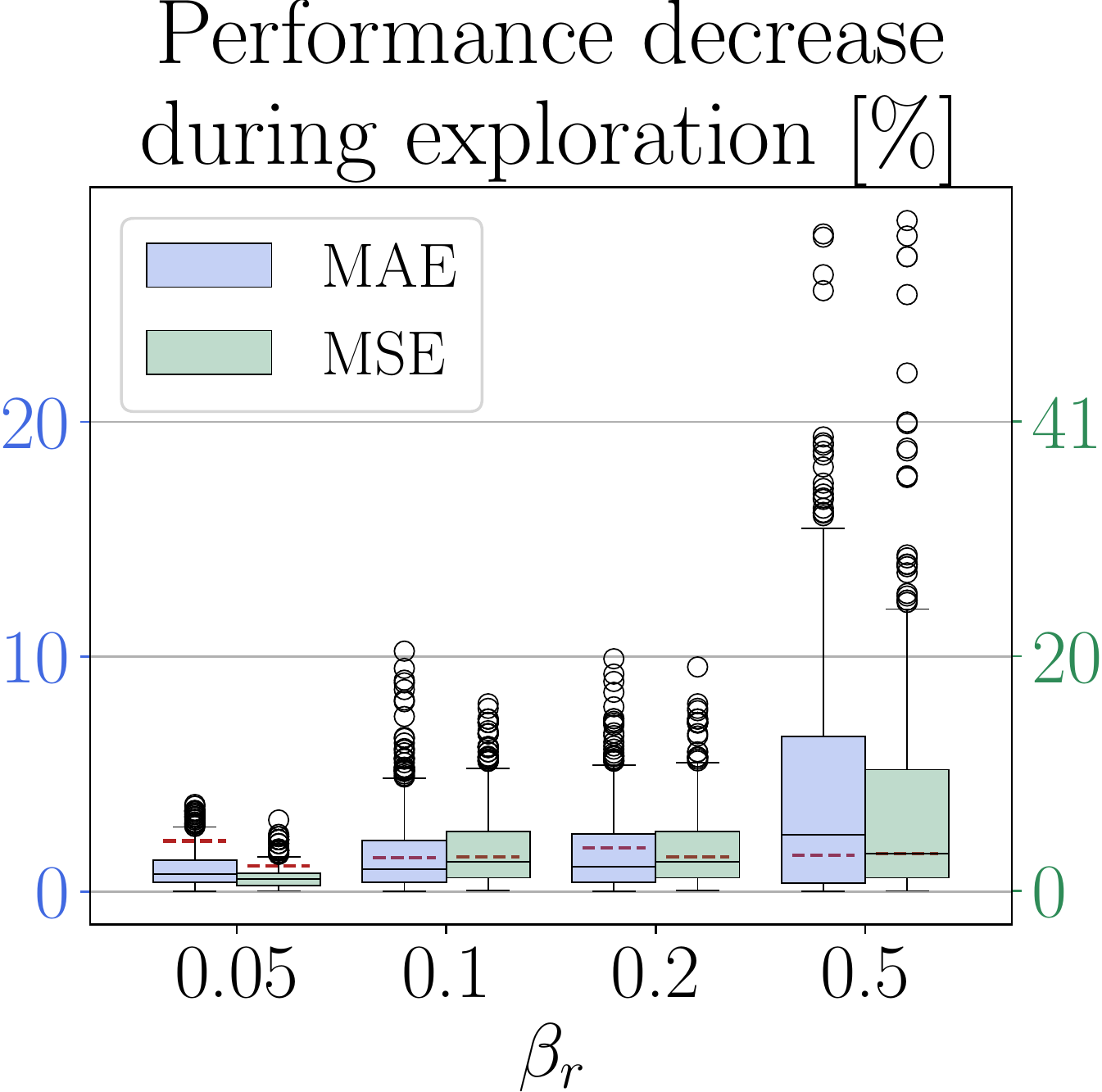}
		\vspace*{-.3cm}
		\label{fig:error_relT}
	\end{subfigure}
	\caption{Comparison of the performance after convergence and during exploration for different values of $\beta_r$, i.e. the allowed deviation from the base controller's signal caused by the residual controller.}
	\label{fig:res_vs_relT}
	\vspace*{-0.3cm}
\end{figure}

The relative constraints are regulated by the parameter $\beta_r$. Fig. \ref{fig:res_vs_relT} shows the influence of $\beta_r$ by comparing both the convergence and exploration performance of a CRRL controller relative to the allowed deviation. As a larger allowed deviation gives the residual controller more freedom, determining $\beta_r$ is a trade-off between the eventual increase in performance attainable and the decrease possible during exploration.

\subsubsection{CRRL performance compared to base policy performance}

\begin{figure}[t]
	\vspace*{-0.3cm}
	\centering
	\begin{subfigure}{.45\columnwidth}
		\centering
		\includegraphics[width=\linewidth]{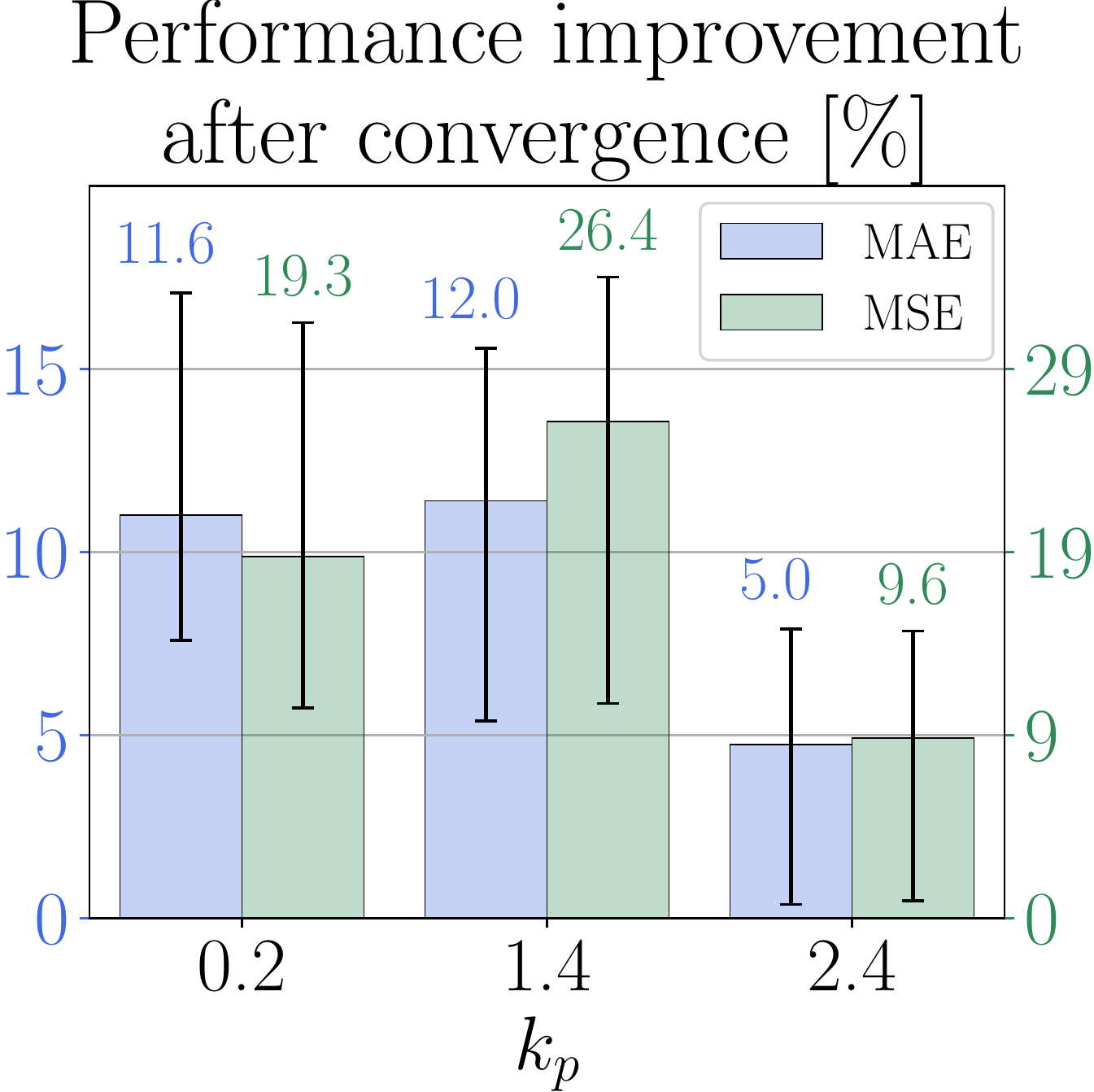}
		\vspace*{-.3cm}
		\label{fig:improv_P}
	\end{subfigure}%
	\begin{subfigure}{.45\columnwidth}
		\centering
		\includegraphics[width=\linewidth]{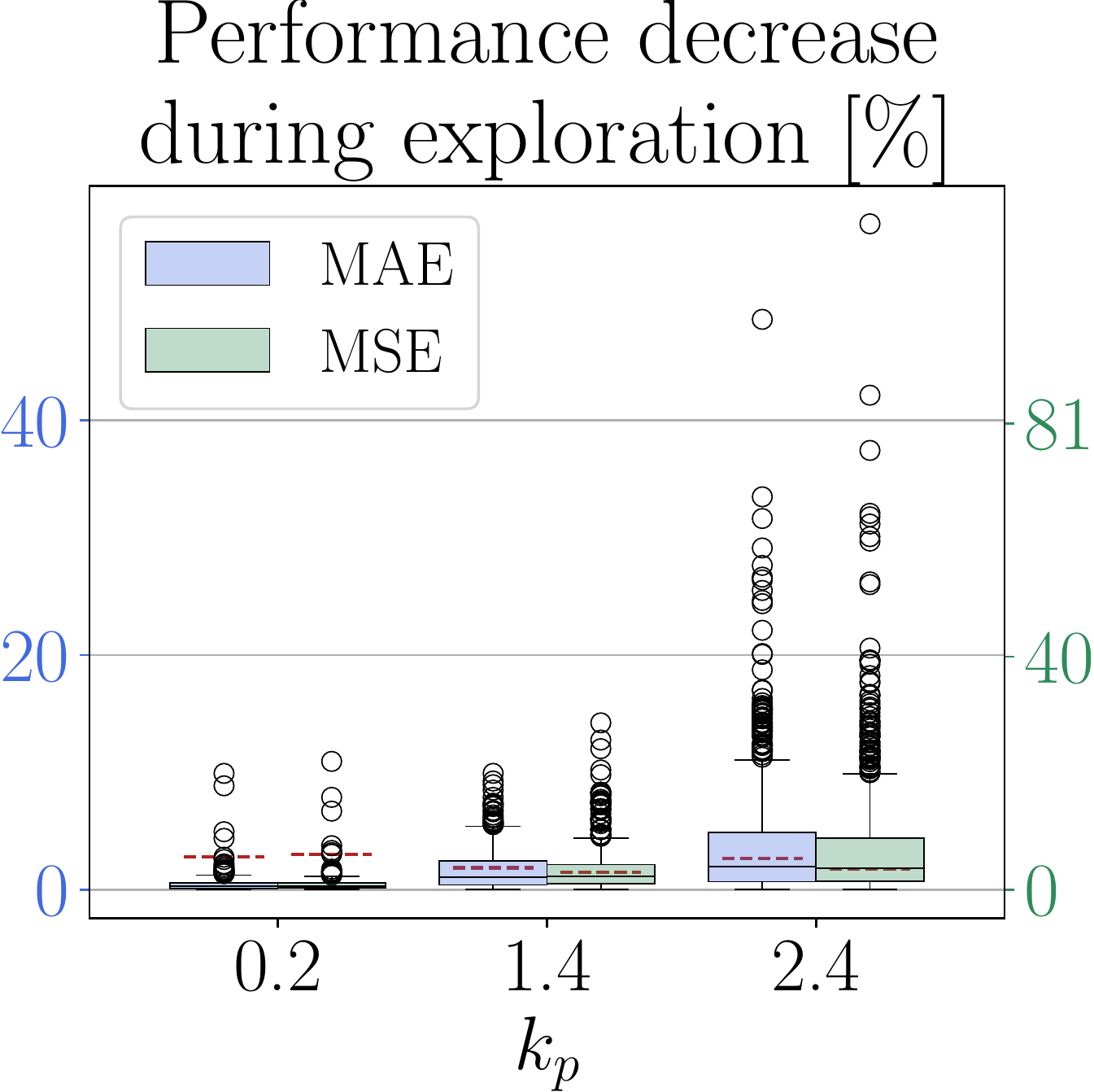}
		\vspace*{-.3cm}
		\label{fig:error_P}
	\end{subfigure}
	\caption{Comparison of the performance after convergence and during exploration for a relative CRRL controller with different base controllers.}
	\label{fig:res_vs_P}
	\vspace*{-.6cm}
\end{figure}

A desirable key feature of CRRL is to not reduce the performance of an already optimal controller. The best PI parameters for the current setup and a constant angular velocity reference of 60 rpm were found to be $k_p = 2.4$ and $k_I = 0.1$. Fig. \ref{fig:res_vs_P} compares both the performance after convergence and during exploration of a CRRL controller relative to the base policy for this PI controller as well as an average ($k_p = 1.4$, $k_I = 0.1$) and a poorly ($k_p = 0.2$, $k_I = 0.1$) tuned one. 

The CRRL controller attains a similar increase in performance for both non-optimally tuned base controllers and it succeeds in improving the best base policy by 5\% on average. Note that the reinforcement learning algorithm sometimes doesn't succeed in finding an improvement for this base controller, in which case it learns to give nearly 0 output to not decrease the performance as the min-max interval line indicates. This result allows to deploy the CRRL methodology on controllers that are likely to be optimal as well, as the residual controller learns to refrain from adapting if no improvement is found, instead of diminishing performance. We also want to emphasize that no additional hyperparameter tuning was carried out to optimize the results for each experiment. For all configurations, the median of the decrease in performance caused by the exploration is less than the maximum decrease that was observed for the base controller itself. Notably for both non-optimally tuned base controllers, the largest observed PI performance decrease is larger than or close to 75\% of the decreases observed after activating the CRRL controller. Although the relative decrease in performance increases as the performance of the base policy increases, larger deviations are seldom and statistical outliers.

\begin{figure}[h]
	\vspace*{-0.5cm}
	\centering
	\begin{subfigure}{.5\columnwidth}
		\centering
		\includegraphics[width=\linewidth]{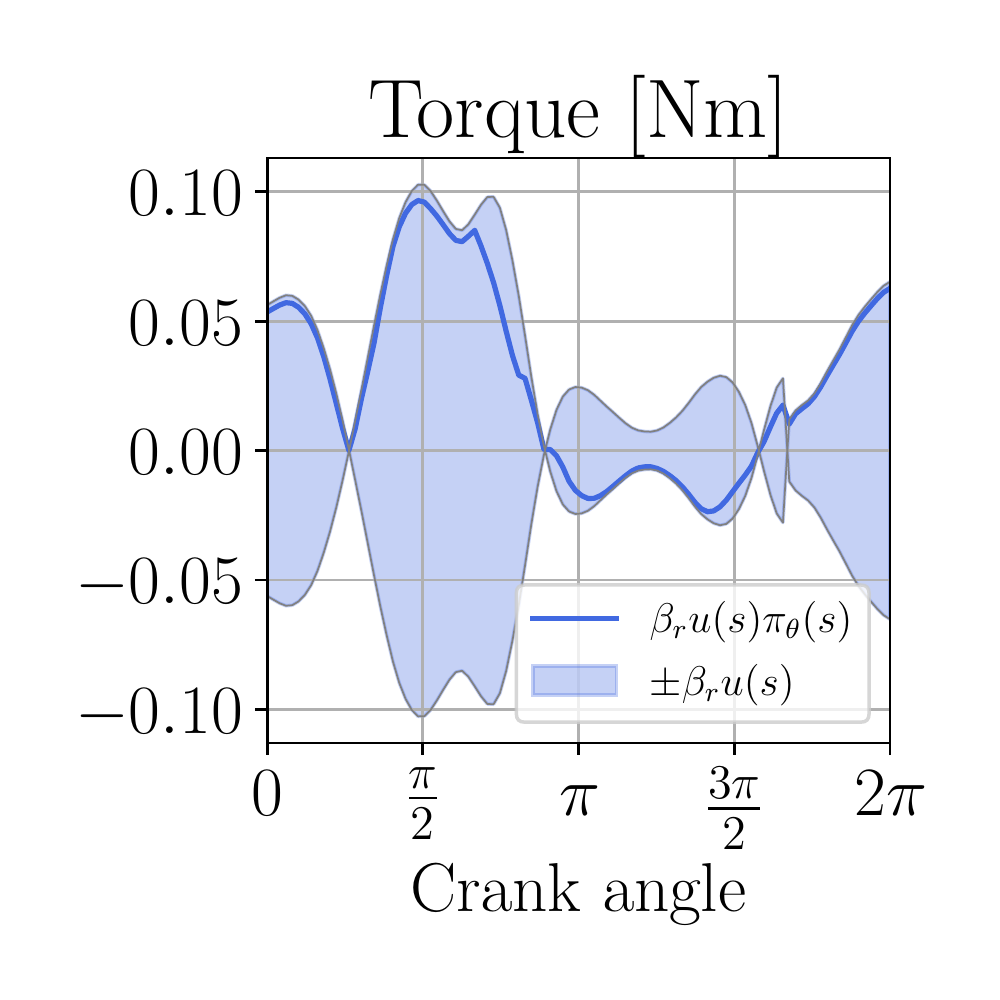}
		\vspace*{-.8cm}
		\label{fig:res_act_ex}
	\end{subfigure}%
	\begin{subfigure}{.5\columnwidth}
		\centering
		\includegraphics[width=\linewidth]{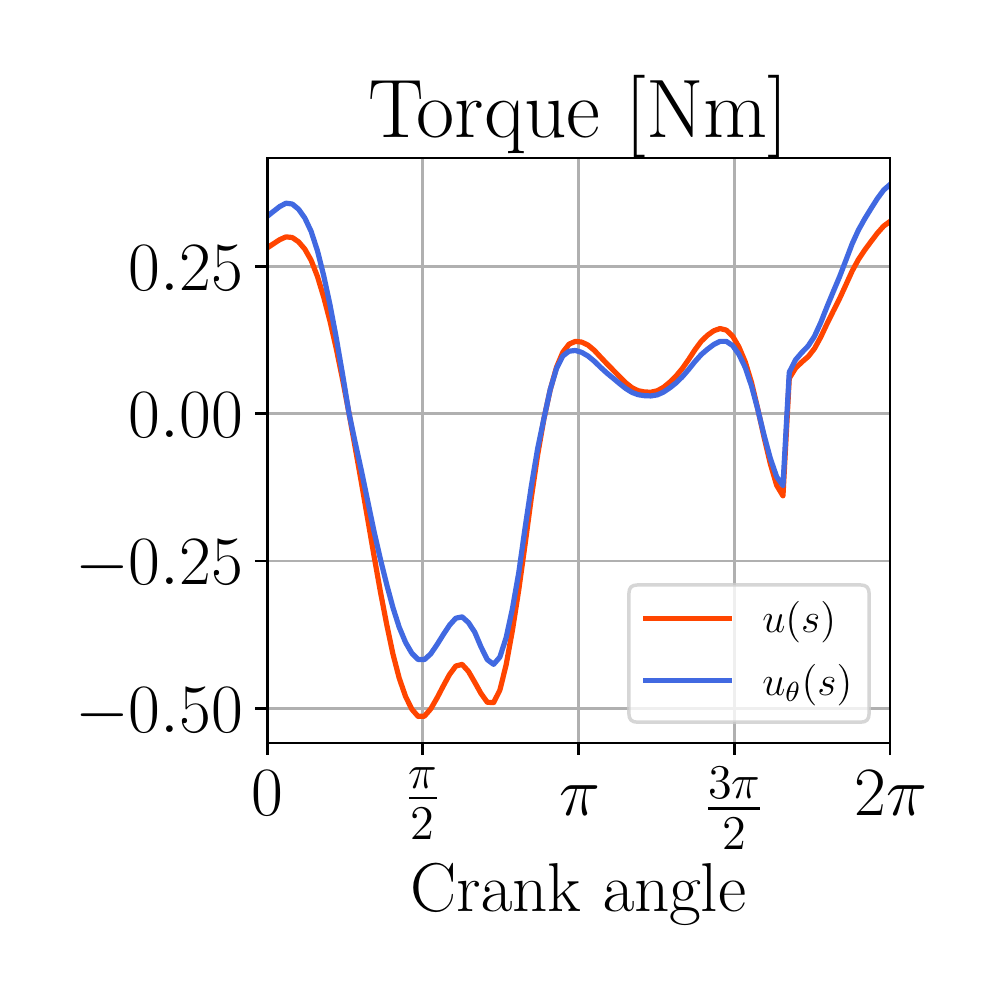}
		\vspace*{-.8cm}
		\label{fig:tot_act_ex}
	\end{subfigure}
	\caption{Left: exemplary residual control signal and its bounds. Right: exemplary base (red) and total control signal (blue).}
	\label{fig:ex_action}
	\vspace*{-0.2cm}
\end{figure}

Fig. \ref{fig:ex_action} on the left shows the residual actions taken by a CRRL controller after convergence with an optimally tuned base PI controller as well as the bounds in between which it is allowed to operate. The residual policy has learned to either add a specific signal, maximally reinforce or maximally counteract the base policy to increase optimality. As the constraints are relative, the residual actions are constrained to 0 at times when the base controller outputs zero as well. The regions where the residual signal reaches the limits of its allowed deviation suggest that a looser bound might result in a more optimal policy after convergence. This is a trade-off with the possible performance loss during the exploration phase, as investigated in Subsection \ref{section:constraints}. Fig. \ref{fig:ex_action} on the right illustrates the base policy's control signal and the total control signal with the residual added. Note the limited changes to the base signal that result in a 5\% performance improvement. 

\subsubsection{CRRL adaptivity to different references}

\begin{figure}[h!]
	\vspace*{-0.3cm}
	\centering
	\begin{subfigure}{.45\columnwidth}
		\centering
		\includegraphics[width=\linewidth]{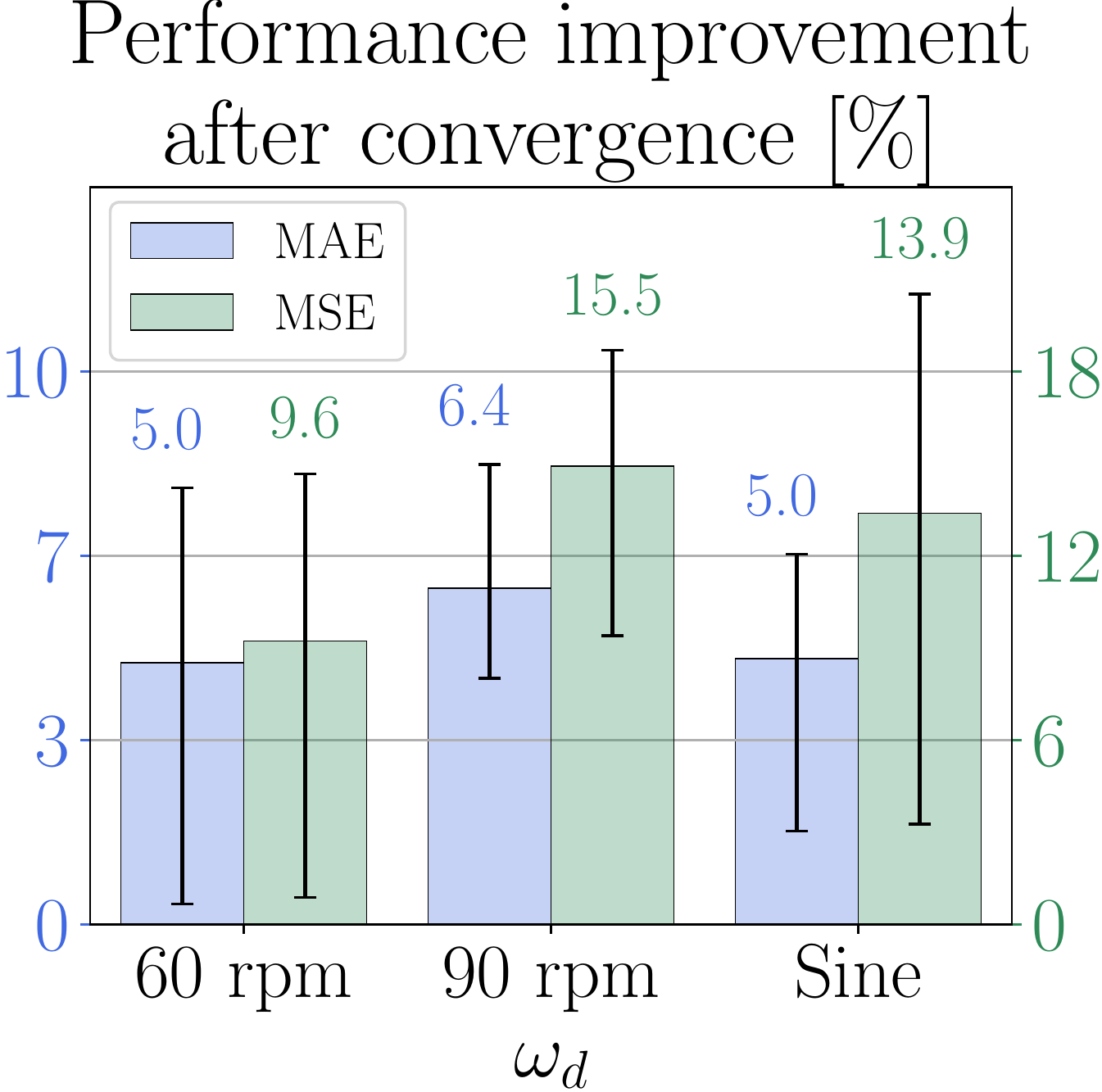}
		\vspace*{-.3cm}
		\label{fig:improv_setpoints}
	\end{subfigure}%
	\begin{subfigure}{.45\columnwidth}
		\centering
		\includegraphics[width=\linewidth]{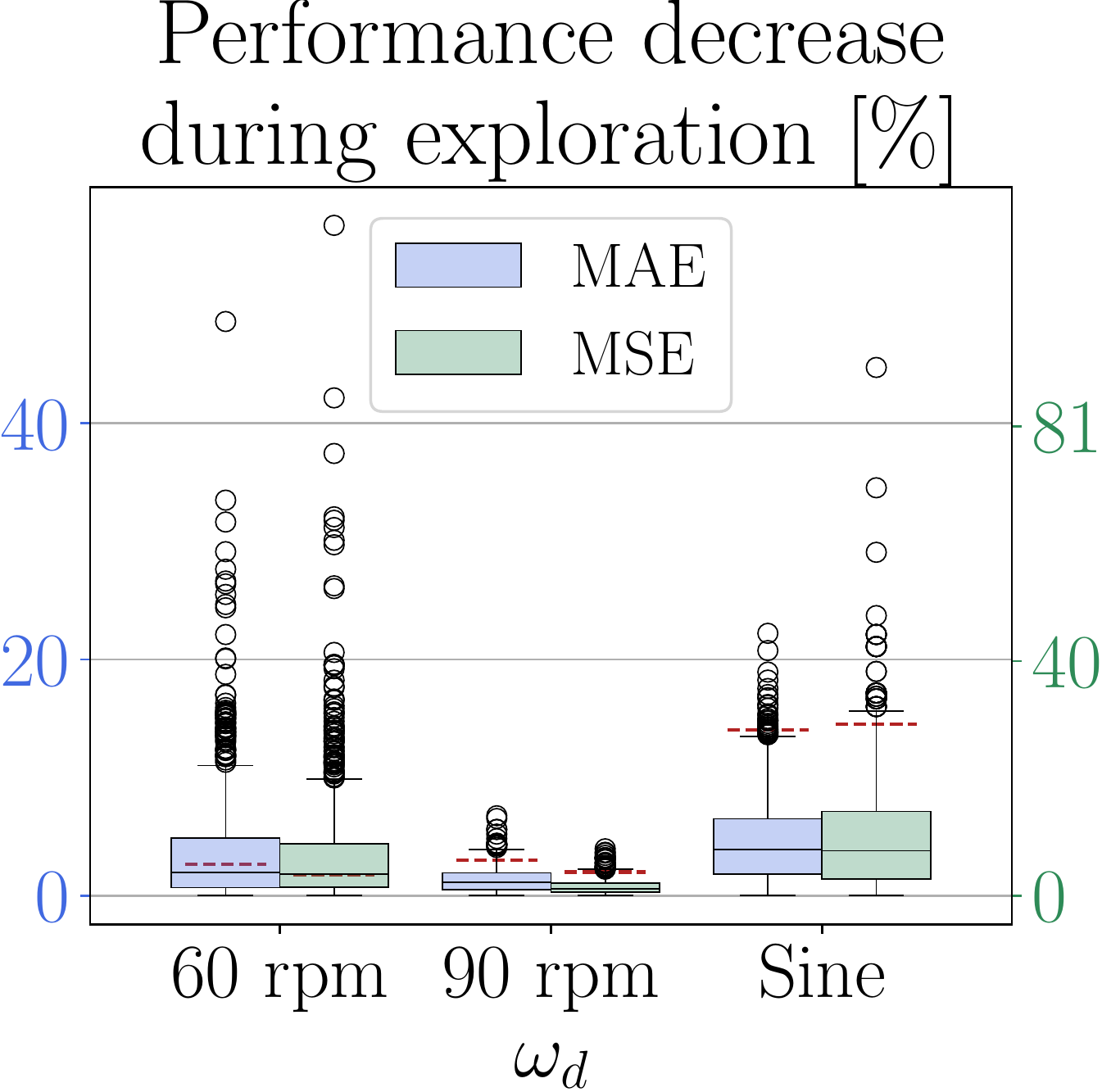}
		\vspace*{-.3cm}
		\label{fig:box_setpoints}
	\end{subfigure}
	\caption{Comparison of the performance after convergence and during exploration for a relative CRRL controller tracking different references.}
	\label{fig:setpoints}
	\vspace*{-.5cm}
\end{figure}

In Fig. \ref{fig:setpoints}, the performance is shown when tracking either a constant reference of 60 or 90 rpm or a sine reference in function of the crank angle, $\omega_d = 15\sin(\psi) + 60$ [rpm], with $\psi$ the crank angle. The base controllers are the ones that were found to have the best performance for each reference, respectively ($k_p = 2.4$, $k_I = 0.1$), ($k_p = 0.7$, $k_I = 0.1$) and ($k_p = 1.4$, $k_I = 0.1$). Note that for the higher reference speed, excessive shaking of the setup caused by a high $k_p$ gain limits the practically feasible values. The figure shows how the same residual controller succeeds in learning improvements for different base controllers tracking different references, demonstrating the adaptivity to operating conditions. Note that for the sinusoidal reference, which is more challenging for the base PI controller to track, only some outlier cycles over all runs have a larger temporary decrease in performance than the largest decrease observed when only the base controller acts. In line with previous results, the difference in base controller for the constant references causes a large difference in relative decrease in performance observed during the exploration phase.

\vspace*{-.1cm}

\section{Conclusions and Future Work}
In this paper, we proposed CRRL, a method to improve the optimality of conventional controllers that are robust, but suffer from suboptimality when faced with uncertain operating conditions. In CRRL, a reinforcement learning algorithm learns corrective adaptations to the conventional controller's output, directly from the controlled system's operating data. By adding the adaptation residually on top of and constraining it by the base controller's output, the robustness of the latter is leveraged to limit the possible performance decrease during the learning process of the residual agent. The Lyapunov stability theory was used to establish safety guarantees of the proposed method even when worst-case conditions are met for a broad class of mechatronic systems. The performance of CRRL was validated experimentally on a slider crank setup tracking a speed reference with a PI base controller. The method is shown to improve the performance for different configurations of the base controller tracking different references substantially after convergence, while maintaining safe operation at all times. The structure of the constraints applied on the residual agent's actions was investigated and it was shown experimentally that constraints relative to the base controller's output are beneficial to limit the possible performance decrease during training while achieving a substantial improvement after convergence. In future work we will focus on expanding the method with adaptive, state-dependent constraints for the residual agent. A consideration is that even though the CRRL architecture limits the performance decrease during exploration greatly, the nearly uniformly sampled actions during exploration inherent to SAC would nonetheless be too brisk for e.g. a position controlled system. As a next step, we will explore how to design the exploration process for different situations to increase the general applicability of CRRL.\\
\vspace*{-.6cm}

\appendices
\section{Tracking stability of mechanical systems}

We start by recollecting some ingredients from Lyapunov Stability theory.

\begin{definition}
	\label{Lyap:quad}
	A strict Lyapunov function $V$ for $\dot{x} = f(x)$ at $x^*$ is quadratic if it is analytic and there exist three positive constants $\alpha_1$, $\alpha_2$ and $\alpha_3$ so that $\alpha_1 \|e\|^2 \leq V \leq \alpha_2 \|e\|^2 \text{ and } \dot V \leq -\alpha_3 \|e\|^2,~ e = x - x^*, ~\forall t\in\mathbb{R}^+$ and $x\in\mathcal{X}$.
\end{definition}

\begin{theorem}
	\label{Lyap:dist}
	Consider a disturbed dynamical system $\dot{x} = {A}(t,x) x + d(t)$. Let $V$ satisfy \cref{Lyap:quad} at $0$ for the undisturbed system (i.e. $d\equiv 0$) on $\mathcal{D}$ and assume there exists a positive constant $\alpha_4$ such that $|\nabla_x V d | \leq \alpha_4 \|x\|$ then the response of the disturbed system from any initial condition $x(t_0)\in\mathcal{D}$ is bounded by $\|x(t)\|\leq e^{-\frac{1}{2} \frac{\alpha_3}{\alpha_2} t} \frac{\alpha_2}{\sqrt{\alpha_1}}\|x(t_0)\|^2 + \frac{\alpha_2 \alpha_4}{\alpha_1 \alpha_3}$.
\end{theorem}
\begin{proof}
	See reference \cite{koditschek1987quadratic}, Theorem 2.
\end{proof}

\vspace*{-.1cm}
Further we are interested in the stability of the closed-loop system dynamics of a mechanical system governed by ${M}(q)\ddot{q} + B(q,\dot{q})\dot{q} + g(q) = u$, $M\succ 0$. We use the feedback linearisation control policy, $u = M(q)\ddot{q}_r + B(q,\dot{q})\dot{q}_r + g(q)  - k_\text{P} \dot{e} - k_\text{I} e$, where $e = q-q_r$ is defined as the tracking error w.r.t. a reference $q_r\in\mathcal{D}$. Further assume that the reference satisfies $\|q_r\|\leq \theta_0$, $\|\dot{q}_r\|\leq\omega_0$, $\|\ddot{q}_r\|\leq \alpha_0$. Finally we assume the system is initialised so that $\|e(0)\|^2 + \|\dot{e}(0)\|^2 \leq \epsilon$ and $\|q(0)\|^2\leq \eta $. The closed-loop system dynamics are given. Note that the dynamics correspond with those in Theorem \ref{Lyap:dist}.
\begin{equation}
\label{system}
\begin{aligned}
\begin{bmatrix}
\dot{e} \\ \ddot{e}
\end{bmatrix} &= \begin{bmatrix}
0 & I \\
- k_\text{I} M^{-1} & - M^{-1}\left(B + k_\text{P} I\right)
\end{bmatrix}\begin{bmatrix}
{e} \\ \dot{e}
\end{bmatrix}
\end{aligned}
\end{equation}

The following theorem allows to make claims about the stability of the system in (\ref{system}).

\begin{theorem}
	\label{th:quad}
	Let $r:\mathbb{R}\mapsto\mathcal{J}$ be a smooth curve, with $\|q_r\|\leq \theta_0 $ and $\|\dot{q}_r\|\leq \omega_0$. Further, 
	let $V:\mathcal{X}\mapsto\mathbb{R}$ be defined as 
	\vspace*{-.2cm}
	\begin{equation}
	V = \tfrac{1}{2} x^\top P(q) x, ~  P(q) = \begin{bmatrix}
	\lambda k_\text{I} I & \frac{1}{2} k_\text{P} I \\
	\frac{1}{2} k_\text{P} I & \lambda M(q)
	\end{bmatrix}
	\vspace*{-.3cm}
	\end{equation}

	with $x=(e,\dot{e})$ and where $\lambda> \frac{k_\text{P}^2}{2k_\text{I}\nu(M)}$, 
	then $V$ is quadratic for system (\ref{system}) with $\alpha_1 = \nu_\eta({P}), ~\alpha_2 = \mu_\eta({P}), ~\alpha_3 = \nu_\eta(\check{Q})$ where $
	\check{Q} = \frac{k_\text{P}}{2 \mu_\eta(M)} \begin{bmatrix}
	\frac{\nu(M)k_\text{I}}{2 \mu_\eta(M)}& \frac{1}{2} k_\text{P} \\
	\frac{1}{2}k_\text{P} & (\lambda-1) \nu(M)
	\end{bmatrix}$
	if $k_\text{I} > \tfrac{\mu_\eta(M)\mu_\eta(B)^2 \omega_0^2}{2\nu(M)^2}$ and $k_\text{P} > \tfrac{2 \mu_\eta(L_q)\omega_0}{1-\frac{1}{\lambda}-\sqrt{\frac{\epsilon^2\mu_\eta(P)\mu_\eta(B)}{\lambda^2 \nu_\eta(P)\nu_\eta(M)}}}$. The norms are defined as $\mu_\eta (A) = \sup_{\|q\|\leq \eta} \sup_{\|x\|= 1} |x^\top A(q) x|$, $\mu(A) = \lim_{\eta \rightarrow \infty} \mu_\eta(A)$ and $\nu_\eta(A) = \inf_{\|q\|\leq \eta } \inf_{\|x\|= 1} x^\top A(q) x$.  Note that from the definition of the norms it follows that $\sup_{\|q\|\leq \eta } x^\top A(q) x \leq \mu_\eta (A) \|x\|^2$ and $\nu_\eta(A)\|x\|^2 \leq \inf_{\|q\|\leq \eta } x^\top A(q) x$.
\end{theorem}
\begin{proof}
	See reference \cite{koditschek1987quadratic}, Proposition 10, Corollary 11 and Proposition 12.
\end{proof}
\vspace*{-.7cm}

\section{Proof of theorem \ref{th:absolute} and \ref{th:relative}}\label{sec:proof-of-theorem-refthabsolute}
\begin{proof}
	We can analyse the tracking stability of an absolute CRRL agent defined as $u =  - k_\text{P} \dot{e} -k_\text{I} e- \beta_a\pi_\theta(q,\dot{q},\dots )$ by analysing the disturbance term
	\vspace*{-0.3cm}
	\begin{equation}
	d = - \begin{bmatrix}
	0 \\
	\ddot{q}_r + M^{-1}\left(B(q,\dot{q}) \dot{q}_r +g(q) +  \beta_a\pi_\theta(q,\dot{q},\dots)\right)
	\end{bmatrix}
	\end{equation}
	and adopting the Lyapunov function from Theorem \ref{th:quad}
\vspace{-2pt}
	{\begin{multline}
		\nabla_x V d = \\ x^\top P \begin{bmatrix}
		0 \\
		\ddot{q}_r + M^{-1}\left(B(q,\dot{q}) \dot{q}_r +g(q) +  \beta_a\pi_\theta(q,\dot{q},\dots)\right)
		\end{bmatrix}\dots \\ \leq \sqrt{\tfrac{1}{4}k_\text{P}^2 + \lambda^2 \mu_\eta(M)^2}
		(\alpha_0 + \tfrac{\mu_\eta(B)\omega_0 + g_0 + \beta_a}{\nu_\eta(M)}) \|x\| 
		\end{multline}}
	\hspace*{-4pt}where $g_0 = \sup \|g(q)\|
	$. Further we rely on the results from Theorem \ref{Lyap:dist}, Definition \ref{Lyap:quad} and Theorem \ref{th:quad} to show that $\|x(t)\| \leq e^{-\frac{1}{2} \frac{\alpha_3}{\alpha_2} t} \frac{\alpha_2}{\sqrt{\alpha_1}} \|x(0)\|^2+\Delta, \lim_{t\rightarrow \infty} \|x(t)\| \leq \Delta$ with $\alpha_1$, $\alpha_2$ and $\alpha_3$ as in Theorem \ref{th:quad}, $\alpha_4$ as defined above and
	\begin{equation}
	\Delta = \tfrac{\mu_\eta({P})\sqrt{\tfrac{1}{4}k_\text{P}^2 + \lambda^2 \mu_\eta(M)^2} \left(\alpha_0 + \tfrac{1}{\nu_\eta(M)}\left(\mu_\eta(B)\omega_0 + g_0 + \beta_a \right)\right)}{\nu_\eta({P})\nu_\eta(\check{Q})}
	\end{equation}
	where $\mu_\eta(B) = \sup_{\|q\|\leq \eta } \sup_{\|x\|= 1} \sup_{\|y\|= 1} \|B(q,x)y\| $. Note that it follows that $\|B(q,x)y\|\leq \mu_\eta (B)\|x\|\|y\|$.
\end{proof}

Analogously we can analyse the tracking stability of a relative CRRL agent defined as $u = -\left( k_\text{P} \dot{e} + k_\text{I} e\right) (1 + \beta_r\pi_\theta(q,\dot{q},\dots ) )$. Here we could either perform a similar analysis as in the previous proof, identifying the associated disturbance and determining the corresponding value for $\alpha_4$. However, this disturbance would depend on the error and may therefore be overly conservative. Alternatively, we could try and analyse the undisturbed closed-loop dynamics corresponding with the control policy $u = M(q)\ddot{q}_r + B(q,\dot{q})\dot{q}_r + g(q) -\left(k_\text{P} \dot{e} + k_\text{I} e\right) (1 + \beta_r\pi_\theta(q,\dot{q},\dots ) )$. Then the closed-loop system dynamics are as in (\ref{system}) but with $k_\text{P}'$ and $k_\text{I}'$ substituted for $k_\text{P}$ and $k_\text{I}$ respectively,  
where $k_\text{P}' = (1+\beta_r\pi_\theta)k_\text{P} $ and  $k_\text{I}' = (1+\beta_r\pi_\theta)k_\text{I} $. If we can derive bounds for $\gamma$, $k_\text{P}$ and $k_\text{I}$ in this context and so that $V$, as defined in \cref{th:quad}, satisfies \cref{Lyap:quad} for the resulting closed-loop system, we can rely again on \cref{Lyap:dist} to establish a bound on the error $\|x\|$.
\vspace*{-.23cm}
\begin{proof}
	The derivative of the Lyapunov function along the motion of the system is given by
\vspace*{-.4cm}
{\begin{multline}
		\dot{V} = - \tfrac{1}{2}k_\text{P} x^\top \begin{bmatrix}
		k_\text{I}' M^{-1} & \tfrac{1}{2}k_\text{P}' M^{-1} \\
		k_\text{I}' M^{-1} & \lambda(1+2\pi_\theta) \\
		\end{bmatrix} x - \tfrac{1}{2}k_\text{P} (\lambda-1)\|\dot{e}\|^2 \\ - \tfrac{1}{2}e^\top M^{-1} B \dot{e} -  \beta_r\pi_\theta k_\text{P}  k_\text{I} e^\top \dot{e} k_\text{P} + \lambda \dot{e}^\top \left(\tfrac{1}{2}\dot{M}-B\right)\dot{e}
\end{multline}}
	For the last term it holds that $\lambda \dot{e}^\top \left(\tfrac{1}{2}\dot{M}-B\right)\dot{e} = \lambda\tfrac{1}{2}\dot{e}^\top L_q(\dot{q}_r) \dot{e} \leq \lambda\mu_\eta(L_q)\omega_0 \|\dot{e}\|^2$	where $
	L_q(x) = \dot{M}(q_x) - \sum_{i=1}^n x_i \partial_{q_i} M
	$
	and $q_x:\mathbb{R}\mapsto\mathcal{D}$ with $\dot{q}_x = x$. After factoring out, $\|\dot{e}\|^2$, for the middle term
	\vspace*{-.4cm}
	{\begin{multline}
		-\tfrac{1}{2}k_\text{P}\|\dot{e}\|^2 \Big(\lambda-1 +  e^\top M^{-1} B(q,\dot{e}) \tfrac{\dot{e}}{\|\dot{e}\|^2} + \dots \\ e^\top M^{-1} B(q,\dot{q}_r) \tfrac{\dot{e}}{\|\dot{e}\|^2} + 2\beta_r\pi_\theta k_\text{I} \tfrac{e^\top \dot{e}}{\|\dot{e}\|^2} \Big) < -\tfrac{1}{2}k_\text{P} \|\dot{e}\|^2 \times \dots \\ \left(\lambda-1-2\beta_r k_\text{I}-\epsilon \sqrt{\tfrac{\mu_\eta(P)\mu_\eta(B)}{\nu_\eta(P)\nu_\eta(M)}}\right) - \tfrac{1}{2}k_\text{P}  e^\top M^{-1} B(q,\dot{q}_r) \dot{e}
		\end{multline}}
	\hspace*{-10pt}since $e^\top M^{-1} B(q,\dot{e}) \tfrac{\dot{e}}{\|\dot{e}\|^2} \leq - \epsilon \sqrt{\tfrac{\mu_\eta(P)\mu_\eta(B)}{\nu_\eta(P)\nu_\eta(M)}} $ and $
	2\beta_r\pi_\theta k_\text{I} \tfrac{e^\top \dot{e}}{\|\dot{e}\|^2} \leq - 2\beta_r k_\text{I}$.
	
	Substituting these inequalities into the equation for $\dot{V}$ yields
	\vspace*{-.4cm}
	{\begin{multline}
		\dot{V} \leq - \tfrac{1}{2}k_\text{P} x^\top \Bigg(\begin{bmatrix}
		k_\text{I}' M^{-1} & \tfrac{1}{2}k_\text{P}' M^{-1} \\
		k_\text{I}' M^{-1} & \lambda(1+2\beta_r\pi_\theta) \\
		\end{bmatrix}-\dots \\ \begin{bmatrix}
		0 & \tfrac{1}{2} M^{-1} B(q,\dot{q}_r) \\
		\tfrac{1}{2} B(q,\dot{q}_r)^\top M^{-1}  & 0 \\
		\end{bmatrix}\Bigg)x - \tfrac{1}{2}k_\text{P} \|\dot{e}\|^\top\times \dots \\ \Bigg(1-\tfrac{1}{\lambda}-\tfrac{2\beta_r k_\text{I}}{\lambda}-\sqrt{\tfrac{\epsilon^2\mu_\eta(P)\mu_\eta(B)}{\lambda^2 \nu_\eta(P)\nu_\eta(M)}} - \tfrac{2\mu_\eta(L_q) \omega_0}{k_\text{P}}\Bigg)
		\end{multline}}

	The second term is $<0$ if $k_\text{P} > \frac{2 \mu_\eta(L_q)\omega_0}{1-\frac{1}{\lambda}-\frac{2\beta_r k_\text{I}}{\lambda}-\sqrt{\frac{\epsilon^2\mu_\eta(P)\mu_\eta(B)}{\lambda^2 \nu_\eta(P)\nu_\eta(M)}}}$. Finally we can rewrite the matrix difference as
	\begin{multline}
	\begin{bmatrix}
	\tfrac{1}{2}\left(k_\text{I}'M^{-1} - \tfrac{1}{2} B^\top M{-2} B\right) & 0\\
	0 & 0 \\
	\end{bmatrix} \\+ \begin{bmatrix}
	\tfrac{1}{2}k_\text{I}' M^{-1} & \tfrac{1}{2}k_\text{P}' M^{-1} \\
	k_\text{I}' M^{-1} & \lambda(1+2\beta_r\pi_\theta) \\
	\end{bmatrix} + \begin{bmatrix}
	\tfrac{1}{2}B^\top M^{-1} \\
	I
	\end{bmatrix} \begin{bmatrix}
	\tfrac{1}{2} M^{-1} B &	I
	\end{bmatrix}
	\end{multline}
	The first matrix is positive semi-definite if $k_\text{I} > \frac{\mu_\eta(M)\mu_\eta(B)^2\omega_0^2}{2(1-\beta_r)\nu(M)^2}$ and the second if $\lambda > \frac{k_\text{P}^2}{2(1-2\beta_r )k_\text{I}\nu(M)}$.
\end{proof}

\vspace*{-.5cm}
\bibliographystyle{Bibliography/IEEEtranTIE}

\bibliography{Bibliography/ResidualReinforcementSlider}
\vspace*{-1.2cm}
\begin{IEEEbiography}[{\includegraphics[width=1in,height=1.25in,clip,keepaspectratio]{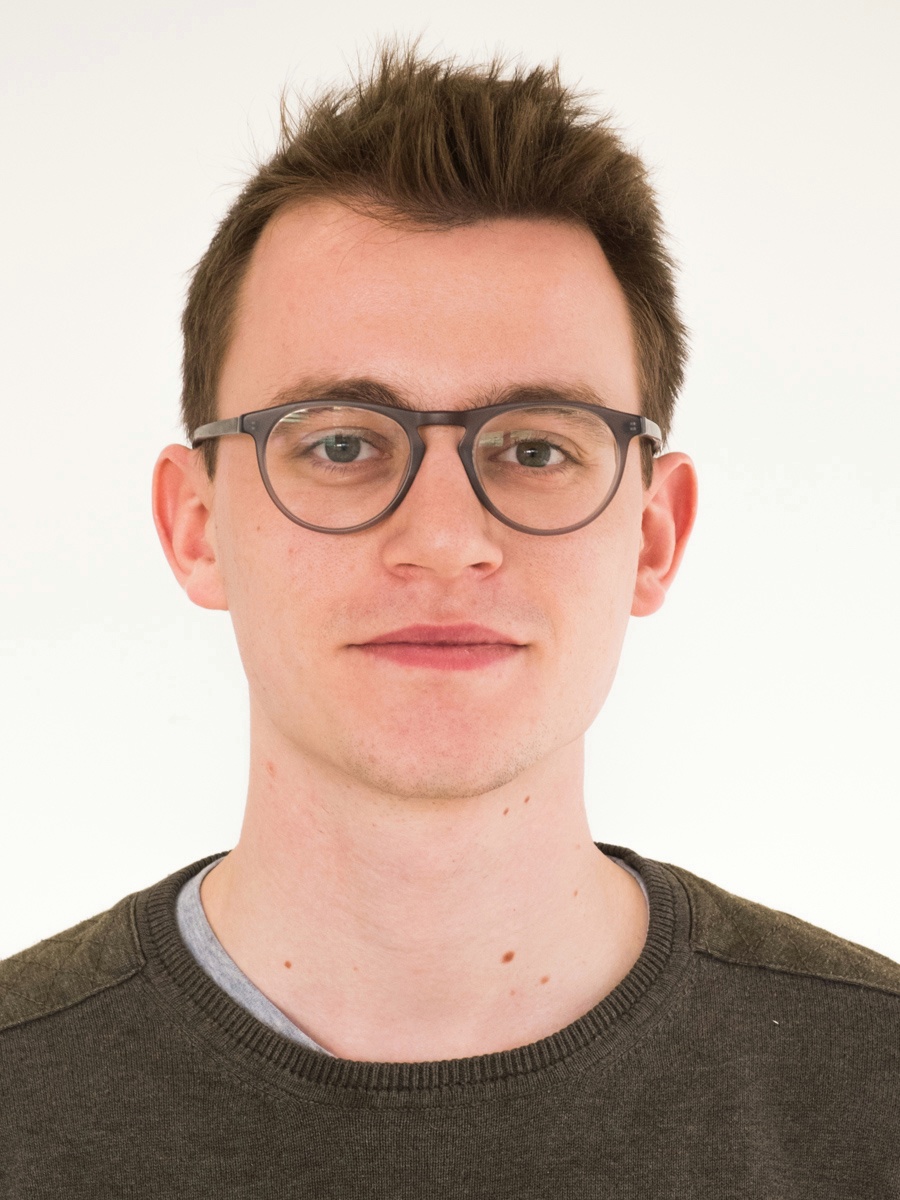}}]
	{Tom Staessens} obtained the M.Sc. in electromechanical and control engineering from Ghent University, Belgium, in 2018. In September 2018, he joined the Department of Electromechanical, Systems and Metal Engineering, where he is currently pursuing the Ph.D. degree. His current research interest includes optimal and safe control by combining traditional control algorithms and data-driven techniques. Tom Staessens is affiliate member of Flanders Make, the strategic research centre for the manufacturing industry in Flanders, Belgium. 
\end{IEEEbiography}
\vspace*{-1.2cm}
\begin{IEEEbiography}[{\includegraphics[width=1in,height=1.25in,clip,keepaspectratio]{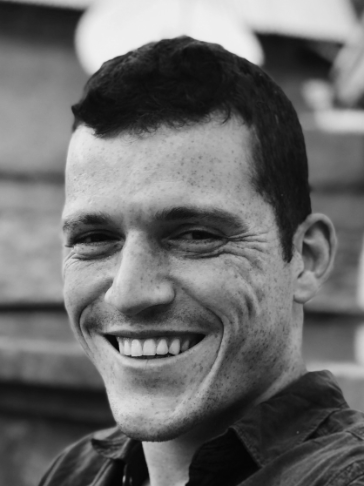}}]
	{Tom Lefebvre} received the M.Sc. in electromechanical and control engineering from Ghent University, Belgium, in 2015. From 2016 till 2019 he pursued the Ph.D. degree at the same University. Since 2019 he is a postdoctoral research assistant. His main research interest includes foundational work on numerical methods for stochastic optimal control, gradient and stochastic trajectory optimization, and uncertainty quantification. Tom Lefebvre is affiliate member of Flanders Make, the strategic research centre for the manufacturing industry in Flanders, Belgium. 
\end{IEEEbiography}
\vspace{-1.2cm}
\begin{IEEEbiography}[{\includegraphics[width=1in,height=1.25in,clip,keepaspectratio]{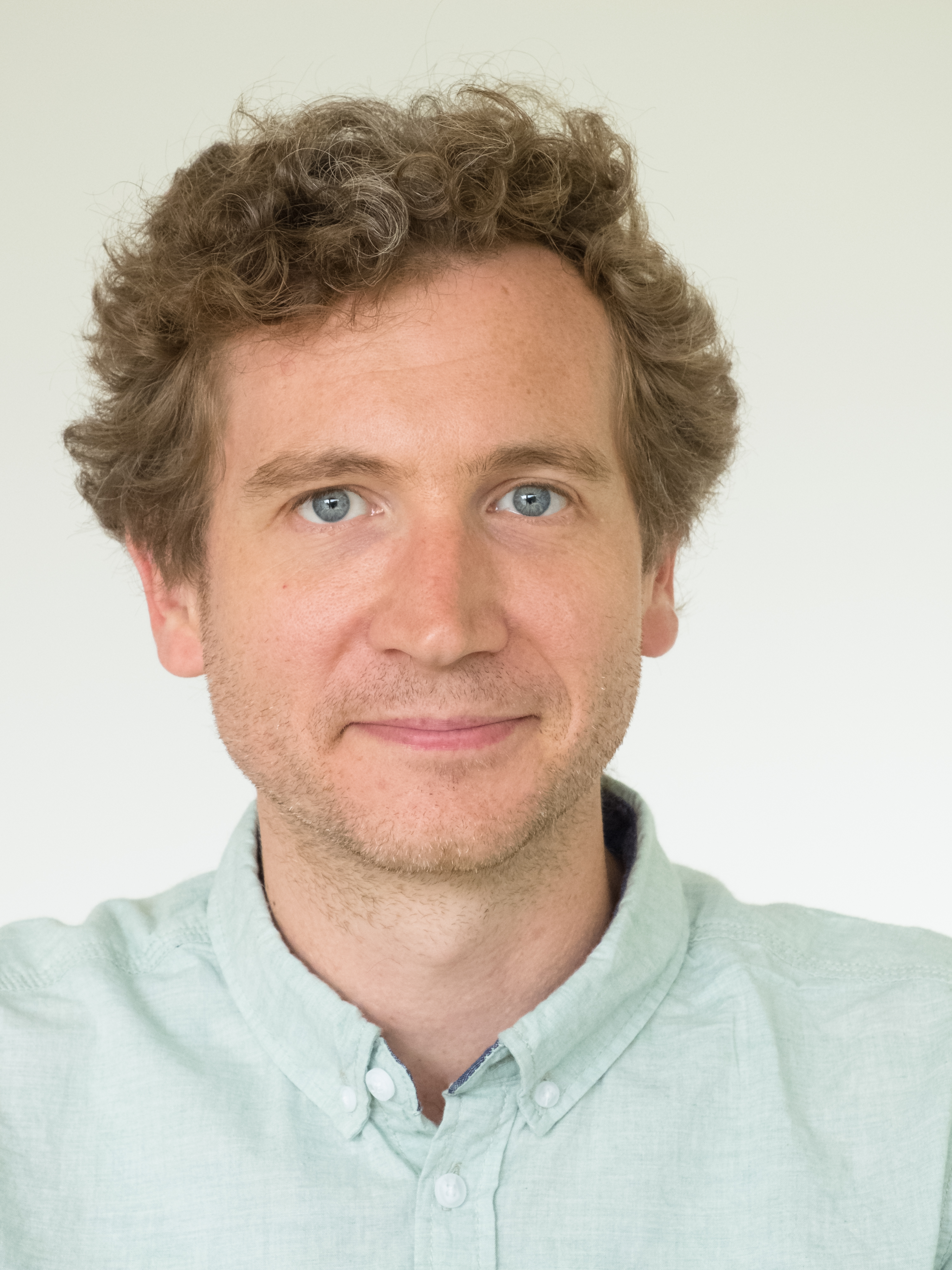}}]
	{Guillaume Crevecoeur} received the MSc and the Ph.D. degree in Engineering Physics from Ghent University in 2004 and 2009, respectively. In 2009 he became a postdoctoral fellow of the Research Foundation Flanders and in 2014 he was appointed Associate Professor at the Faculty of Engineering and Architecture of Ghent University. He is active member of Flanders Make, the strategic research center for the manufacturing industry. His research interests are the modelling, optimization and control of mechatronic and industrial robotic applications.
	
\end{IEEEbiography}

\end{document}